\documentclass[journal]{IEEEtran}
\usepackage{newtxtext,newtxmath}
\usepackage{amsmath,amsfonts}
\usepackage{algorithmic}
\usepackage{algorithm}
\usepackage{array}
\usepackage{enumitem}
\usepackage[caption=false,font=footnotesize,labelfont=sf,textfont=sf]{subfig}
\usepackage{textcomp}
\usepackage{stfloats}
\usepackage{xcolor}
\usepackage{url}
\usepackage{verbatim}
\usepackage{graphicx}
\graphicspath{{figs/}}
\usepackage{cite}

\ifCLASSINFOpdf
 
\else

\fi

\begin{document}

\title{Information-Epidemic Dynamics in Cyber-Physical Systems: A Hypergraph Framework with Interpersonal Relationships}

\author{Shanchao Peng, Minyu Feng, Senior Member, IEEE, Liang-Jian Deng, Senior Member, IEEE, Matja\v{z} Perc, Member, IEEE, and J{\"u}rgen Kurths
\thanks{S.P. and M.F. were supported by the Natural Science Foundation of Chongqing under Grant CSTB2025YITP-QCRCX0007 and the National Natural Science Foundation of China (NSFC) under Grant 62206230. M.P. was supported by the Slovenian Research and Innovation Agency under Grant P1-0403. \textit{(Corresponding author: Minyu Feng.)}

Shanchao Peng and Minyu Feng are with the College of Artificial Intelligence, Southwest University, Chongqing 400715, China (e-mail: myfeng@swu.edu.cn).

Liang-jian Deng is with the School of Mathematical Sciences, University of Electronic Science and Technology of China, Chengdu 611731, China.

Matja\v{z}~Perc is with the Faculty of Natural Sciences and Mathematics, 
University of Maribor, 2000 Maribor, Slovenia, also with the Community Healthcare Center Dr. Adolf Drolc Maribor, 2000 Maribor, Slovenia, also with the Department of Physics, Kyung Hee University, Dongdaemun, Seoul 02447, Republic of Korea, 
and also with the University College, Korea University, Seongbuk-gu, Seoul 02841, Republic of Korea.

J{\"u}rgen Kurths is with the Department of Complexity Science, Potsdam Institute for Climate Impact Research, 14473 Potsdam, Germany, and also with the Department of Physics, Humboldt University of Berlin, 12489 Berlin, Germany.
}
}

\maketitle

\begin{abstract}
Understanding how information propagation affects epidemic dynamics has become an emerging topic of interest. However, the influence of interpersonal relationship heterogeneity on information acquisition and disease transmission has been largely overlooked. In this work, we introduce a hypergraph structure for Cyber-Physical Systems (CPSs) with two distinct layers. The upper layer, referred to as the cyber layer, consists of a mixed hypergraph, capturing both pairwise propagation and higher-order diffusion of epidemic-related information. The lower layer, referred to as the physical layer, employs a Susceptible--Infected--Susceptible (SIS) process to capture epidemic spreading. This work introduces an adaptive perception-protection mechanism based on Jaccard similarity, which accounts for interpersonal heterogeneity. In this mechanism, individuals receive information based on their relationships with neighbors and take protective measures accordingly. We analyze the impact of interpersonal relationships and the adoption of neighborhood-based self-protection strategies on epidemic dynamics. Furthermore, we conduct a theoretical analysis based on the Microscopic Markov Chain Approach (MMCA), analytically derive the outbreak threshold, and confirm the results with extensive Monte Carlo (MC) simulations. The results show that stronger interpersonal relationships can promote information propagation, significantly increase the threshold for epidemic outbreaks, and effectively suppress the scale of the epidemic. The study provides theoretical support for designing epidemic control strategies considering interpersonal heterogeneity and improves the understanding of epidemic spreading on hypergraphs.
\end{abstract}


\IEEEpeerreviewmaketitle

\section{Introduction}
\label{sec1:introduction}
\IEEEPARstart{W}{ith} the continuous development of society and the acceleration of globalization, infectious diseases pose a severe threat to public health and social stability. From the historical Black Death epidemic to SARS and the recent COVID-19 pandemic~\cite{gaudart2010demographic},~\cite{dye2003modeling},~\cite{cheng2024sustainable}, the frequent outbreaks and widespread impact of infectious diseases have sparked widespread concern about their prevention and control. Ye \textit{et al.} presented a fully integrated system for contact tracing of infectious diseases using COVID-19 as a case study~\cite{9858073}. To understand the mechanisms of epidemic spreading in depth, numerous studies have been dedicated to the construction and optimization of epidemiological models, attempting to reveal their evolutionary patterns and provide theoretical support for the formulation of public health policies. 
Duncan J. Watts \textit{et al.} first proposed the WS small-world network model~\cite{watts1998collective}, followed by Albert-L\'aszl\'o Barab\'asi \textit{et al.}, who introduced the BA scale-free network model~\cite{barabasi1999emergence}. These models serve as a basis for studying epidemic dynamics on complex networks. In such networks, nodes represent individuals in the real world, while edges represent interactions or connections between them~\cite{strogatz2001exploring}. Subsequently, in order to capture the dynamics of epidemic spreading in real-world scenarios with greater precision, several classical epidemic models have been developed~\cite{grassly2008mathematical}, including the SIS model~\cite{gray2011stochastic} and the Susceptible-Infected-Recovered (SIR) model~\cite{zaman2008stability}. As the understanding of epidemic dynamics has deepened, increasingly refined models have been developed. Li \textit{et al.} introduced individual protection levels as random variables into the Susceptible-Infected-Recovered-Susceptible (SIRS) model, demonstrating that such heterogeneity significantly affects epidemic spreading~\cite{li2021protection}. Zhou \textit{et al.} considered the problem of detecting a single rumor source from observed network snapshots under the susceptible-exposed-infected-recovered (SEIR) model~\cite{8682104}. Furthermore, Chen \textit{et al.} developed a Susceptible-Infectious-Quarantine-Recovered-Susceptible (SIQRS) model that incorporates isolation measures and studied its dynamics on simplicial complexes~\cite{chen2024siqrs}. Following this, Li \textit{et al.} constructed a metapopulation model based on the MMCA to simulate SIS dynamics within structured communities~\cite{li2025epidemic}. Overall, these epidemic models offer a framework for understanding the mechanisms underlying epidemic dynamics.

While the above epidemic models provide valuable insights into disease dynamics, they often overlook that real-world systems are inherently composed of multiple interacting networks. For example, Lee \textit{et al.} applied the concept of multiplex networks to capture non-additive effects of multiple types of interactions~\cite{lee2014multiplex}. Building upon this, Salehi \textit{et al.} provided a review of the main models, results, and applications of multilayer spreading processes and discussed some promising research directions~\cite{salehi2015spreading}. Subsequently, De Domenico \textit{et al.} reviewed recent advances in understanding spreading processes on multilayer networks and highlighted a variety of physical phenomena that emerge from their complex structures~\cite{de2016physics}. When an infectious disease emerges in a population, behavioral responses to the outbreak can significantly influence its progression. Specifically, upon becoming aware of an epidemic, individuals may proactively wear masks and reduce social contact to lower their risk of infection~\cite{gao2022epidemic}. Funk \textit{et al.} integrated awareness diffusion into epidemiological models and demonstrated that, in well-mixed populations, awareness spreading can reduce outbreak sizes without altering the epidemic threshold~\cite{funk2009spread}. Based on previous research, Granell et al.~\cite{granell2013dynamical} introduced multiplex networks, which led to the rise of CPS. In the CPS, the upper cyber layer models the dissemination of epidemic-related information, while the lower physical layer captures the epidemic dynamics, with a one-to-one correspondence between the nodes in both layers. Subsequent studies have increasingly employed CPS to describe the coupling between awareness diffusion and epidemic spreading. For example, Feng \textit{et al.} developed a two-layer network model that includes individuals who ignore the epidemic, and found that individuals with significant centrality in the awareness layer have a considerable impact on mitigating the spread of infectious diseases~\cite{feng2023impact}. Mishkovski \textit{et al.} demonstrated that the threshold for epidemic outbreaks in CPS may be higher than in isolated networks~\cite{mishkovski2017interplay}, which shows that in CPS, the dissemination of epidemic-related information can effectively suppress the outbreak of the disease. Salehi \textit{et al.} discovered that spreading processes, such as information propagation among users of online social networks and the diffusion of pathogens among individuals through contact networks, are fundamental phenomena in these networks~\cite{salehi2015spreading}. Zhuang \textit{et al.} indicated that understanding the dynamics of information propagation among humans calls for a multi-layer network model where an online social network is conjoined with a physical network~\cite{7542505}. Wang \textit{et al.} studied information diffusion within the CPSS network~\cite{9403368}. The application of CPS in engineering and industry is also becoming increasingly widespread. Su \textit{et al.} focused on the node recovery optimization of cyber-physical power systems and found that using a scoring-based method for node recovery optimization can significantly improve the reliability and power supply capability of the system~\cite{su2024node}. Shang \textit{et al.} proposed a new multistage adversarial game model to protect smart grids from cyber-physical attacks~\cite{shang2025multistage}. Du \textit{et al.} introduced an island-maximizing attack mechanism to maximize the number of power islands in cyber-physical power systems by disrupting lines~\cite{11196948}. Meanwhile, the rapid development of CPS has attracted increasing research attention to CPS security. Yu \textit{et al.} provided a comprehensive view of the security of CPSs from three perspectives, summarizing current security threats and state-of-the-art defense mechanisms to offer researchers a systematic understanding of the field~\cite{10163904}. Building on these insights, recent studies have further advanced the understanding of information-epidemic coupling by incorporating richer behavioral and informational mechanisms. Yang \textit{et al.} investigated the influence of competing public attention on epidemic dynamics and showed that non-pandemic ``hot topics'' may distract collective attention from epidemic-related information, thereby weakening protective responses~\cite{yang2023information}. Their study established a dual-layer MMCA-based framework that links public attention dynamics with epidemic spreading using empirical indicators such as online attention data and epidemic incidence statistics. Building on this idea, Xu \textit{et al.} introduced information asymmetry into multilayer epidemic-information models and demonstrated that unequal access to reliable information can generate heterogeneous protective behaviors among individuals~\cite{xu2025asymmetric}. In addition, Liu \textit{et al.} investigated the global stability and optimal control of epidemic systems with adaptive behavioral responses, and derived theoretical conditions for the global convergence of epidemic dynamics in heterogeneous populations~\cite{liu2023global}. Although these studies progressively enrich behavioral mechanisms in information-epidemic coupling—ranging from attention dilution to information asymmetry and adaptive behavioral responses—they predominantly rely on traditional pairwise networks, which fail to capture group-level interactions commonly observed in real-world Internet of Things systems and social systems.

\begin{figure*}[!t]
\centering
\includegraphics[width=0.95\textwidth]{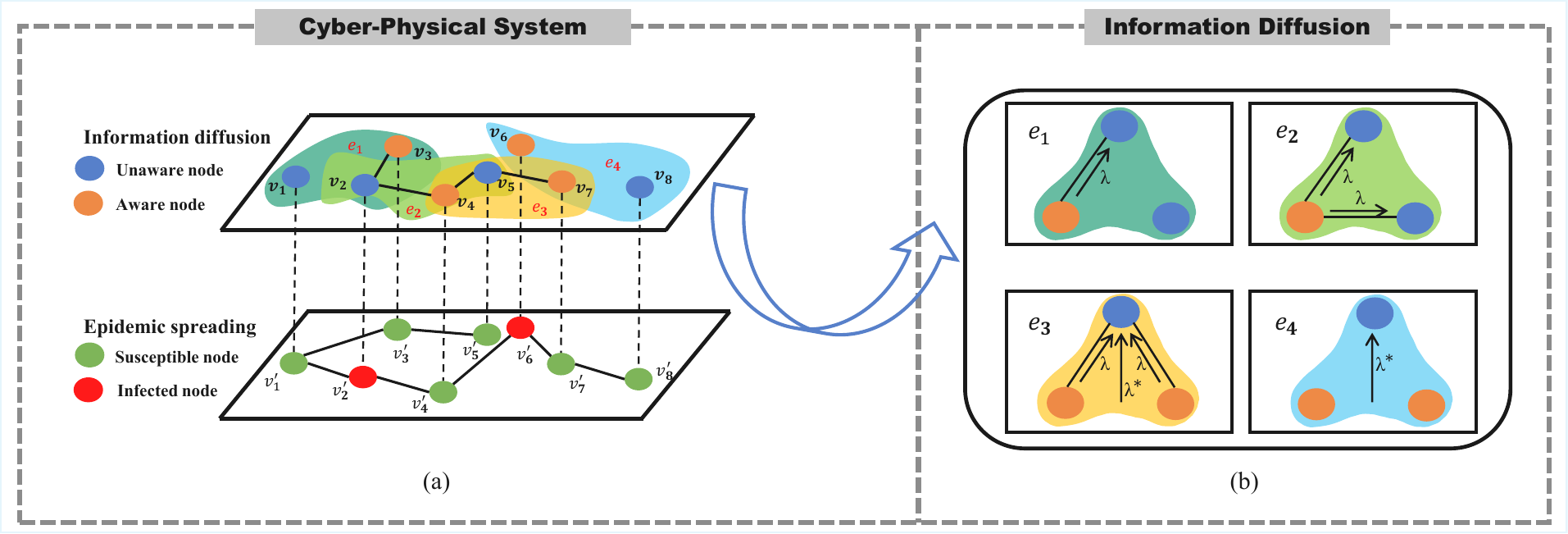}
\caption{Proposed CPS and schematic of information diffusion. (a) Coupling between the cyber and physical layers(left). The upper layer represents the cyber network, modeling the diffusion of epidemic-related information. Each node exists in one of two states: unaware (blue) or aware (orange). Edges between nodes indicate pairwise information transmission, while shaded regions (e.g., $e_1$, $e_2$, $e_3$, and $e_4$) correspond to hyperedges that facilitate higher-order information diffusion. The lower layer denotes the physical contact network, where nodes are either susceptible (green) or infected (red). Edges in this layer denote physical interactions through which the disease spreads. For clarity, we denote nodes in the cyber layer as $v_i$, while nodes in the physical layer are denoted as $v_i'$. Each pair $(v_i, v_i')$ corresponds to the same individual participating simultaneously in both processes. Dashed lines indicate the one-to-one correspondence between nodes in the cyber and physical layers. The entire system is modeled as an undirected and unweighted composite multilayer network. (b) Information diffusion process on the cyber layer(right). 
In $e_1$ and $e_2$, pairwise information transmission occurs between $U$ and $A$ nodes with probability~$\lambda$. However, since only one node within the hyperedge is in the $A$ state, the condition for higher-order information propagation is not satisfied. In $e_4$, this condition is met, and higher-order information transfer occurs within the hyperedge with probability~$\lambda^{*}$. $e_3$ illustrates a scenario in which both pairwise and higher-order information transmissions occur simultaneously, with probabilities~$\lambda$ and~$\lambda^{*}$, respectively. Additionally, at each time step, an aware node may forget the information and revert to the unaware state with probability~$\delta$.
}
\label{fig_1}
\end{figure*}

In real-world scenarios, information interactions often involve more than two individuals, such as group chats on social media, meetings, or classroom discussions. Such multi-participant interactions cannot be fully captured by traditional pairwise network models~\cite{yan2024multisensor}. To address this limitation, Wang \textit{et al.} reformulated the network system as a simplicial complex, introducing a novel framework for social communication that captures not only interactions between individual nodes but also higher-order interactions among groups of individuals~\cite{wang2020social}. Building on this idea, Yuan \textit{et al.} introduced physical-layer information as a new source of awareness and found that it can effectively suppress epidemic spreading~\cite{yuan2025impacts}. Zhang \textit{et al.} developed a new framework of hypergraph signal processing (HGSP) based on the tensor representation to generalize the traditional graph signal processing (GSP) to tackle high-order interactions~\cite{8887197}. Chen \textit{et al.} proposed a composite effective degree Markov chain approach (CEDMA) to describe the discrete-time epidemic dynamics on higher-order networks~\cite{chen2023composite}. Moreover, the results reported in~\cite{wang2024epidemic} indicate that incorporating a 2-simplex structure in the information layer can effectively increase the epidemic threshold and reduce the infection scale.

As mentioned above, although various epidemic spreading models and information diffusion patterns have been widely studied, their application to real-world scenarios still faces several key limitations. (i) While interactions involving three or more individuals are common in practice, most existing models capture group interactions solely through a 2-simplex structure~\cite{iacopini2019simplicial}, with limited exploration of frameworks that can fully represent higher-order interactions. (ii) Many current studies assume that individuals receive and respond to epidemic information uniformly~\cite{li2023coevolution, nie2022markovian}, while neglecting the heterogeneity in individuals' sensitivity and perception. To address these limitations of existing work, we propose a hypergraph structure for CPS in this paper. Our system introduces an adaptive perceptual protection mechanism based on the Jaccard similarity, which accounts for the heterogeneity of interpersonal relationships. Extensive simulations demonstrate that the theoretical prediction results obtained using the MMCA are highly consistent with those from Monte Carlo (MC) simulations, thereby confirming the accuracy of our model.

In summary, the key contributions of this work are as follows.

\begin{enumerate}[label=\arabic*)]
\item We propose a mixed hypergraph structure for CPS. The cyber layer consists of a hypergraph composed of multiple nodes and hyperedges. Nodes within the same hyperedge represent individuals belonging to a common social group. Pairwise interactions between nodes capture direct information propagation, while hyperedges describe the diffusion of higher-order information within the group. The physical layer simulates the spread of the epidemic through real-world physical contacts.

\item We introduce an adaptive perceptual protection mechanism based on the Jaccard similarity, which accounts for the heterogeneity of interpersonal relationships. In this mechanism, the heterogeneity of individuals' relationships with their neighbors directly influences both the amount and strength of information they receive, thereby determining the likelihood of adopting protective measures.

\item We use the MMCA to derive the threshold for epidemic outbreaks in the CPS analytically and verify the theoretical results through extensive MC simulations. Our framework captures the interaction mechanism between information diffusion and disease spread, highlighting the moderating effect of information diffusion on the dynamic evolution of epidemics.
\end{enumerate}

The structure of the paper is outlined as follows. In Sec.~\ref{sec2:model description}, we introduce the construction of the nonlinear cyber-physical networked system and a perception-driven adaptive protection mechanism. In Sec.~\ref{sec3:state evolution and epidemic threshold analysis based on MMCA}, we employ the MMCA to analyze the state changes of each node and analytically derive the epidemic outbreak threshold. In Sec.~\ref{sec4:simulations}, we validate the effectiveness of the model through simulations and analyze the impact of interpersonal relationships on epidemic dynamics. Finally, in Sec.~\ref{sec5:conclusion and outlook}, we provide a comprehensive summary of this study and propose potential directions for future research.

\section{Model Description}
\label{sec2:model description}
In this section, driven by the challenge to characterize group-based information diffusion and its impact on epidemic transmission, we develop a hypergraph model for nonlinear cyber-physical networked systems. Furthermore, we propose a perception-driven adaptive protection mechanism to capture individuals' behavioral responses to epidemic information.

\subsection{Hypergraph Modeling for Cyber-Physical Systems}
We constructed a mixed hypergraph model for CPS to investigate the coupled dynamics of information diffusion and epidemic spread in groups. 
The cyber layer captures information diffusion, while the physical layer simulates disease transmission through physical contact. To capture the characteristics of real-world group-based information exchange, we build the cyber layer as a mixed hypergraph $H = (V, E, E_p)$ where $V = \{v_1, v_2, \dots, v_n\}$ is the set of nodes, $E = \{e_1, e_2, \dots, e_m\}$ is the set of hyperedges, and $E_p = \{(v_i, v_j), (v_i, v_k), \dots, (v_j, v_k)\}$ is the set of pairwise edges. This is an extension of a traditional graph, where each hyperedge \( e_j \) is a nonempty subset of the node set \( V \). 
In our model, we adopt a mixed hypergraph, where nodes within the same hyperedge represent individuals belonging to the same social group. This structure allows us to simultaneously consider pairwise information interactions and higher-order information diffusion induced by hyperedges, while avoiding introducing additional topological complexity. Since our model primarily investigates higher-order information diffusion, sensory feedback mechanisms, and their coupling with epidemic spreading, we aim to maintain a physical-layer network with moderate structural complexity and without excessive degree heterogeneity. Prior studies indicate that, when the average degree is comparable, WS and BA networks produce qualitatively consistent dynamical outcomes. These outcomes include the presence of phase transitions and the role of feedback in promoting or suppressing spreading. The main differences lie in the numerical values of the corresponding epidemic thresholds, rather than the qualitative characteristics of the transmission mechanisms themselves. For these reasons, we use a WS network to model the physical layer, because it captures two essential characteristics of real contact networks, namely local clustering and short average path length, while preserving analytical tractability and avoiding the high level of heterogeneity present in scale-free networks. This modeling choice allows us to more clearly separate the influence of the coupling mechanism between information dissemination and epidemic spread, and ensures that the principal conclusions of our study remain generalizable across a wider range of network topologies. We represent the system as undirected and unweighted, as shown in Fig.~1(a), where a one-to-one correspondence is established between the nodes in the two layers. In this way, each individual is present in both the cyber and the physical layers, indicating that each individual is affected by both information diffusion and epidemic spread. 

Individuals actively acquire epidemic-related information through interactions with their neighbors and may gradually forget it over time. In the cyber layer, we adopt the UAU model, where U and A represent the unaware and aware states, respectively. An individual transitions from the unaware to the aware state through three pathways: (i) upon infection, an individual becomes aware immediately; (ii) pairwise interactions with aware neighbors make the individual aware with the probability $\lambda$; and (iii) higher-order interactions within hyperedges transmit awareness to the individual with the probability $\lambda^{*}$. Notably, disease-related information propagates to an unaware node through higher-order interactions only when two aware nodes simultaneously exist in the same hyperedge. Additionally, as time goes by, aware nodes may forget the epidemic information and return to the unaware state with probability $\delta$. Fig.~1(b) illustrates these information propagation mechanisms, which actively shape the evolution of awareness across the network. In the physical layer, we implement the SIS model, where S and I denote the susceptible and infected states, respectively. A susceptible individual becomes infected with probability $\beta$ when contacting an infected neighbor, while an infected individual recovers to the susceptible state with probability $\mu$.

To construct an information propagation network with group-based interactions, we design a method for generating a connected mixed hypergraph based on a random growth mechanism. The construction process of the cyber layer is illustrated in Algorithm~\ref{alg:hypergraph}.

As shown in Algorithm~\ref{alg:hypergraph}, the proposed procedure constructs a connected mixed hypergraph in four phases. In the initialization phase, three distinct nodes are randomly selected to form the first hyperedge, ensuring that the hypergraph is nonempty. These nodes are added to the covered set \(C\), which contains the nodes that have been assigned to one or more hyperedges during the hypergraph construction. During the incremental construction phase, an anchor node from the covered set is combined with two nodes from the uncovered set \(R\) — the set of nodes not yet included in any hyperedge — to iteratively expand the hyperedge set. Here, \(|E|\) denotes the current number of hyperedges in the hypergraph, and the process continues until \(|E|\) reaches the predefined total number of hyperedges. In the connectivity ensuring phase, any isolated nodes are connected to the existing structure by forming new hyperedges with two covered nodes, ensuring full connectivity. After constructing the hyperedge set \(E\), pairwise edges are added to represent dyadic interactions within each hyperedge. For every hyperedge \( e = \{v_a, v_b, v_c\} \in E \), all node pairs \((v_x, v_y)\) are examined, and a pairwise edge is added to \(E_p\) with probability \(p_1\). This process enables the mixed hypergraph to capture both pairwise and higher-order interactions.
In summary, this construction produces a uniform and fully connected mixed hypergraph, which effectively captures group-based interactions and serves as the underlying structure for simulating information diffusion in the proposed UAU-SIS spreading model.

\begin{algorithm}[t]
\caption{Constructing a Mixed Hypergraph}
\label{alg:hypergraph}
\begin{algorithmic}[1]
\REQUIRE Node set \(V\); predefined number of hyperedges \(M\); pairwise connection probability \(p_1\)
\ENSURE A connected mixed hypergraph \(H = (V, E, E_p)\)

\STATE \textbf{Initialization:}
\STATE Randomly select three distinct nodes \(v_i, v_j, v_k \in V\)
\STATE Form the first hyperedge \(E = \{\{v_i, v_j, v_k\}\}\)
\STATE Initialize the covered node set \(C = \{v_i, v_j, v_k\}\)
\STATE Initialize the uncovered node set \(R = V \setminus C\) 

\STATE \textbf{Incremental Construction:}
\WHILE{$|E| < M$}
  \STATE Randomly select an anchor node \(v_{\text{anchor}} \in C\)
  \STATE Randomly select two uncovered nodes \(v_{u_1}, v_{u_2} \in R\)
  \STATE Form a new hyperedge \(\{v_{\text{anchor}}, v_{u_1}, v_{u_2}\}\)
  \STATE Update \(E\), \(C\), and \(R\)
\ENDWHILE

\STATE \textbf{Ensure Hypergraph Connectivity:}
\FOR{each isolated node \(v_{\text{iso}} \in R\)}
  \STATE Randomly select two nodes \(v_{c_1}, v_{c_2} \in C\)
  \STATE Form a hyperedge \(\{v_{\text{iso}}, v_{c_1}, v_{c_2}\}\)
  \STATE Update \(E\) and \(C\)
\ENDFOR

\STATE \textbf{Pairwise Connection:}
\STATE Initialize pairwise edge set \(E_p = \emptyset\)
\FOR{each hyperedge \(e = \{v_a, v_b, v_c\} \in E\)}
  \FOR{each node pair \((v_x, v_y)\) in \(e\)}
    \IF{a random number \(r \in [0,1)\) satisfies \(r < p_1\)}
      \STATE Add a pairwise edge \((v_x, v_y)\) to \(E_p\)
    \ENDIF
  \ENDFOR
\ENDFOR

\RETURN \(H = (V, E, E_p)\)
\end{algorithmic}
\end{algorithm}

\subsection{Adaptive Perception-Protection Mechanism}
In real life, individuals perceive information related to the epidemic and then take corresponding protective measures. When constructing the proposed CPS, we consider the impact of interpersonal closeness on the information reception process in the cyber layer. To capture the strength of the relationship between individuals, we adopt the Jaccard similarity, which effectively reflects the extent of shared social contexts between two individuals. In hypergraph-based interactions, a higher overlap of hyperedges indicates that two individuals frequently participate in the same groups or activities, leading to stronger mutual perception and a higher likelihood of information exchange. Specifically, for any two nodes \(v_i\) and \( v_j\) that appear in at least one common hyperedge, the closeness of their relationship $W_{v_i v_j}$ is defined as:
\begin{equation}
\label{eq:1}
W_{v_i v_j} = \left( \frac{|E(v_i) \cap E(v_j)|}{|E(v_i) \cup E(v_j)|} \right)^{\alpha},
\end{equation}
where \( E(\cdot) \) denotes the set of hyperedges containing a given node. $W_{v_i v_j}$ represents the $\alpha$-th power of the ratio between the intersection and union sizes of the hyperedge sets for the nodes $v_i$ and $v_j$, and its value lies in the interval $[0,1]$. The regulation factor $\alpha$ adjusts the closeness between nodes $v_i$ and $v_j$, capturing the nonlinear variation in their relationship. When $\alpha = 0$, $W_{v_i v_j} = 1$ for all node pairs, indicating uniform relationship strength across the network with no differentiation between individuals. $W_{v_i v_j}$ quantifies the extent to which two nodes share hyperedges in the hypergraph. Higher values indicate that the nodes participate together in more groups, reflecting stronger interpersonal relationships. In real-world scenarios, this corresponds to individuals frequently interacting in multiple social, professional, or interest-related groups. 

Based on~\eqref{eq:1}, the total closeness of relationships between node \( v_i \) and all its neighbors is defined as:
\begin{equation}
\label{eq2}
W_{v_i} = \sum_{v_j \in N(v_i)} W_{v_i v_j},
\end{equation}
where \( N(v_i) \) represents the set of neighbors of node \( v_i \).

When node $v_i$ receives information from different neighbors, its degree of acceptance varies heterogeneously. To quantitatively characterize the differentiated acceptance strength of node $v_i$ towards information transmitted by each neighbor, we define the attenuation factor $\gamma_{v_i v_j}$ as follows:
\begin{equation}
\label{eq:3}
\gamma_{v_i v_j} = \left( \frac{W_{v_i v_j}}{W_{v_i}} \right)^{\eta}.
\end{equation}

In real life, individuals can reduce their infection risk by adopting self-protection measures, such as wearing masks or reducing social outings. In this context, $\gamma_{v_i v_j}$ serves as an attenuation factor that quantifies the influence of these protective behaviors on lowering the effective infection rate. The regulation factor $\eta \geq 0$ is used to capture the nonlinear responsiveness of individuals and controls the strength of the effect of self-protection mechanisms on disease transmission. As $\eta$ increases, $\gamma_{v_i v_j}$ decreases, indicating a stronger inhibitory effect of awareness on the spread of the epidemic. Notably, when $\eta = 0$, the cyber layer exerts no influence on the epidemic dynamics, and the model reduces to a baseline SIS spreading process without behavioral response. 

To distinguish between aware and unaware nodes, we denote $\beta^U$ as the infection probability of an unaware susceptible individual, and $\beta^A$ as that of an aware susceptible individual. In addition, when the node is in the unaware state, \( \beta^U \) is numerically equivalent to \( \beta \). Therefore, the relationship can be expressed as:
\begin{equation}
\label{eq4}
\beta^A_{v_i v_j} = \gamma_{v_i v_j} \cdot \beta^U = \left( \frac{W_{v_i v_j}}{W_{v_i}} \right)^{\eta} \cdot \beta.
\end{equation}

\section{State Evolution and Epidemic Threshold Analysis Based on MMCA}
\label{sec3:state evolution and epidemic threshold analysis based on MMCA}
In this section, we analyze the state changes of each node and analytically derive the epidemic threshold of the proposed CPS based on the MMCA dynamic equations.

\begin{figure}[!t]
\centering
\includegraphics[width=0.47\textwidth]{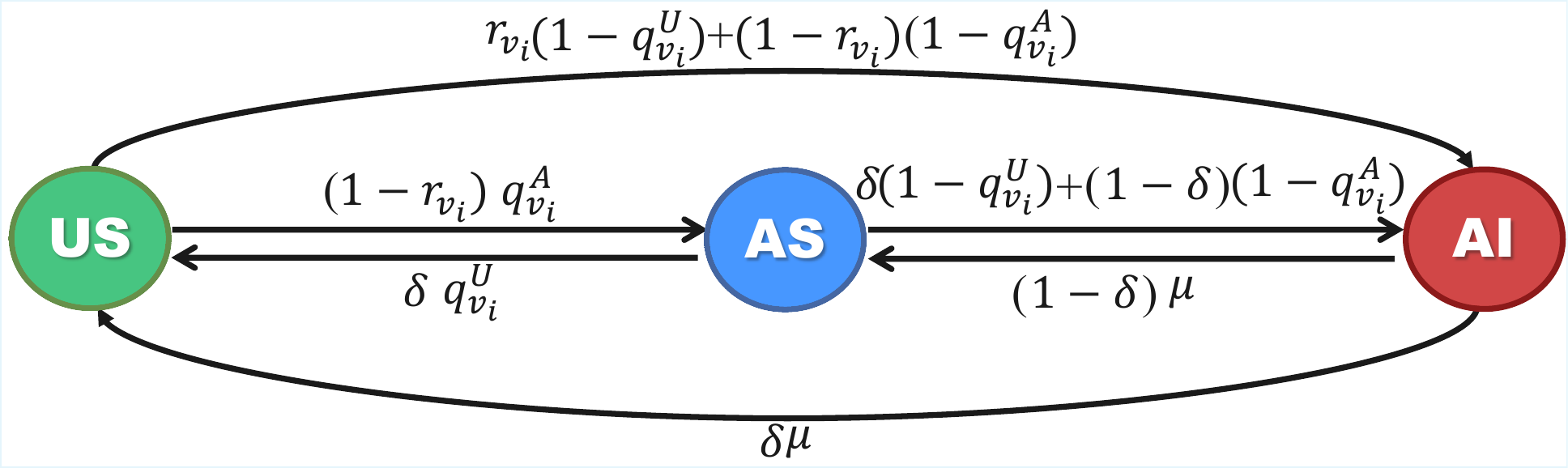}
\caption{State transitions of the UAU-SIS model on the CPS. The probability that a node $v_i$ in the aware state (A) escapes infection from all its infected neighbors is denoted by $q^A_{v_i}$, while $q^U_{v_i}$ represents the probability that a node in the unaware state (U) avoids infection. Let $r_{v_i}$ represent the probability that a node receives no information from any of its aware neighbors. The parameter $\delta$ denotes the probability that an aware individual forgets the epidemic-related information, while $\mu$ describes the probability that an infected (I) individual recovers and returns to the susceptible (S) state.}
\label{fig_2}
\end{figure}

\subsection{State Evolution Analysis via MMCA}
To investigate the co-evolutionary dynamics of epidemic spreading and information diffusion in the CPS, we developed a coupled UAU--SIS model based on the MMCA framework.

In our model, we assume that individuals become aware immediately upon infection, so the state of being unaware and infected (\( \mathrm{UI} \)) does not exist. This assumption is introduced to simplify the state space of the coupled UAU--SIS dynamics and to facilitate the theoretical analysis based on the MMCA framework. From a practical perspective, this assumption can be interpreted as a scenario in which infected individuals rapidly recognize symptoms or are quickly identified through testing or health monitoring systems, thereby becoming aware of their infection status. Introducing unaware infected individuals would delay the awareness feedback mechanism and may weaken the suppression effect of awareness on epidemic spreading. However, the fundamental coupling mechanism between information diffusion and disease transmission remains unchanged. Therefore, a node can reside in one of the following three conditions: unaware and susceptible (\( \mathrm{US} \)), aware and susceptible (\( \mathrm{AS} \)), or aware and infected (\( \mathrm{AI} \)). At time step \( t \), the probabilities of node \( v_i \) being in the three possible states—US, AS, and AI—are denoted by \( p^{US}_{v_i}(t) \), \( p^{AS}_{v_i}(t) \), and \( p^{AI}_{v_i}(t) \), respectively. Furthermore, these probabilities follow the normalization condition \( p^{US}_{v_i}(t) + p^{AS}_{v_i}(t) + p^{AI}_{v_i}(t) = 1 \). Let \( a_{v_i v_j} \) and \( b_{v_i v_j} \) represent the elements of the adjacency matrices for the cyber layer and the physical layer, respectively. The variable \( c_{v_i v_j v_k} \) indicates whether the nodes \( v_i \), \( v_j \), and \( v_k \) within the same hyperedge follow the conditions for higher-order information propagation. If the condition holds, \( c_{v_i v_j v_k} = 1 \); otherwise, \( c_{v_i v_j v_k} = 0 \). We define \( r_{v_i}(t) \) as the probability that node \( v_i \) fails to receive information from any aware neighbor at time step \( t \), expressed as \( r_{v_i}(t) = r^{(1)}_{v_i}(t) \cdot r^{(2)}_{v_i}(t) \). Here, \( r^{(1)}_{v_i}(t) \) denotes the probability that node \( v_i \) fails to receive information through pairwise communication with its neighbors, while \( r^{(2)}_{v_i}(t) \) represents the probability of failure in obtaining information via higher-order interactions. It is worth emphasizing that higher-order interactions capture a group-driven reinforcement effect that cannot be decomposed into pairwise interpersonal similarity. For this reason, the higher-order term $r_{v_i}^{(2)}$ is modeled using a constant probability~$\lambda^{*}$. In addition, \( q^U_{v_i}(t) \) and \( q^A_{v_i}(t) \) denote the probabilities that node \( v_i \) is not infected by any of its infected neighbors when in the US and AS states, respectively. Then we have:
\begin{equation}
\label{eq:5}
\left\{
\begin{aligned}
q^A_{v_i}(t) &= \prod\nolimits_{v_j} \left[1-b_{v_jv_i}p^{AI}_{v_j}(t)\beta^A\right], \\
q^U_{v_i}(t) &= \prod\nolimits_{v_j} \left[1-b_{v_jv_i}p^{AI}_{v_j}(t)\beta^U\right], \\
r^{(1)}_{v_i}(t) &= \prod\nolimits_{v_j} \left[1-a_{v_jv_i}p^A_{v_j}(t)\lambda W_{v_iv_j}\right], \\
r^{(2)}_{v_i}(t) &= \prod\nolimits_{v_j,v_k} \left[1-c_{v_iv_jv_k}p^A_{v_j}(t)p^A_{v_k}(t)\lambda^*\right], \\
r_{v_i}(t) &= r^{(1)}_{v_i}(t)r^{(2)}_{v_i}(t),
\end{aligned}
\right.
\end{equation}
where $p^A_{v_j}(t) = p^{AI}_{v_j}(t) + p^{AS}_{v_j}(t)$. Fig.~2 illustrates the transition process of each state. We derive the coupled evolution equations of the system using the MMCA. For each node $v_i$, the following holds:
\begin{equation}
\label{eq:6}
\left\{
\begin{array}{l}
p^{US}_{v_i}(t+1) = p^{US}_{v_i}(t) \, r_{v_i}(t) \, q^U_{v_i}(t) \\[5pt]
\qquad \qquad \quad \quad + p^{AS}_{v_i}(t) \, \delta \, q^U_{v_i}(t) \\[5pt]
\qquad \qquad \quad \quad + p^{AI}_{v_i}(t) \, \delta \, \mu, \\[8pt]

p^{AS}_{v_i}(t+1) = p^{US}_{v_i}(t) \left[1 - r_{v_i}(t)\right] q^A_{v_i}(t) \\[5pt]
\qquad \qquad \quad \quad + p^{AS}_{v_i}(t) (1 - \delta) q^A_{v_i}(t) \\[5pt]
\qquad \qquad \quad \quad + p^{AI}_{v_i}(t) (1 - \delta) \mu, \\[8pt]

p^{AI}_{v_i}(t+1) = 
p^{US}_{v_i}(t) \left(1 - r_{v_i}(t)\right) \left[1 - q^A_{v_i}(t)\right] \\[5pt]
\qquad \qquad \quad \quad + p^{US}_{v_i}(t) r_{v_i}(t) \left[1 - q^U_{v_i}(t)\right] \\[5pt]
\qquad \qquad \quad \quad + p^{AS}_{v_i}(t) \delta \left[1 - q^U_{v_i}(t)\right] \\[5pt]
\qquad \qquad \quad \quad + p^{AS}_{v_i}(t) (1 - \delta) \left[1 - q^A_{v_i}(t)\right] \\[5pt]
\qquad \qquad \quad \quad + p^{AI}_{v_i}(t) \left[ \delta (1 - \mu) + (1 - \delta)(1 - \mu) \right].
\end{array}
\right.
\end{equation}

\subsection{Epidemic Threshold Analysis}
\label{subsec:threshold}
When the information diffusion and epidemic spreading processes reach a steady state in the model, we obtain:
\begin{equation}
\begin{cases}
p^{US}_{v_i}(t+1) = p^{US}_{v_i}(t) = p^{US}_{v_i}, \\[5pt]
p^{AS}_{v_i}(t+1) = p^{AS}_{v_i}(t) = p^{AS}_{v_i}, \\[5pt]
p^{AI}_{v_i}(t+1) = p^{AI}_{v_i}(t) = p^{AI}_{v_i}.
\end{cases}
\end{equation}

Near the epidemic outbreak threshold, the infection probability of each node is assumed to be very small. Therefore, the MMCA equations can be linearized around the disease-free equilibrium by neglecting higher-order terms of the infection probability. Under this approximation, the derived analytical threshold characterizes the onset of epidemic outbreaks. However, it does not capture the full nonlinear dynamics of the system far from the critical point. Consequently, we assume that $p^{AI}_{v_i} = \epsilon_{v_i}$, where $\epsilon_{v_i} \to 0^{+}$. Under this assumption, the first two equations in~\eqref{eq:5} can be rewritten as follows:
\begin{equation}
\label{eq:8}
\begin{cases}
q^A_{v_i} \approx 1 - \beta^A \displaystyle\sum_{v_j} b_{v_i v_j} \, \epsilon_{v_j}, \\[8pt]
q^U_{v_i} \approx 1 - \beta^U \displaystyle\sum_{v_j} b_{v_i v_j} \, \epsilon_{v_j}.
\end{cases}
\end{equation}

Substituting~\eqref{eq:8} into~\eqref{eq:6}, we derive:
\begin{equation}
\label{eq:9}  
\begin{cases}
p^{US}_{v_i} = p^{US}_{v_i} \, r_{v_i} + p^{AS}_{v_i} \, \delta, \\[6pt]
p^{AS}_{v_i} = p^{US}_{v_i} \, (1 - r_{v_i}) + p^{AS}_{v_i} \, (1 - \delta), \\[6pt]
\mu \, \epsilon_{v_i} = \left( p^{AS}_{v_i} \, \beta^A + p^{US}_{v_i} \, \beta^U \right) \displaystyle\sum_{v_j} b_{v_j v_i} \, \epsilon_{v_j}.
\end{cases}
\end{equation}

Since $p^{AI}_{v_i} = \epsilon_{v_i} \approx 0$ as $\beta$ approaches the critical outbreak threshold $\beta_c$, we have $p^{A}_{v_i} = p^{AI}_{v_i} + p^{AS}_{v_i} \approx p^{AS}_{v_i}, \quad \text{and} \quad p^{AS}_{v_i} + p^{US}_{v_i} \approx 1.$ Substituting these relations, the third expression in~\eqref{eq:9} is expressed as:
\begin{equation}
\label{eq:10}  
\mu \epsilon_{v_i} = \sum_{v_j} \left( p^{A}_{v_i} \, \gamma_{v_i v_j} \, \beta^U + (1 - p^{A}_{v_i}) \beta^U \right) b_{v_i v_j} \, \epsilon_{v_j}.
\end{equation}

Then, further processing~\eqref{eq:10} leads to:
\begin{equation}
\label{eq:11}  
\sum_{v_j} \left\{ \big[ 1 - (1 - \gamma_{v_i v_j}) p^{A}_{v_i} \big] b_{v_i v_j} - \frac{\mu}{\beta} E_{v_i v_j} \right\} \epsilon_{v_j} = 0,
\end{equation}
where \(E_{v_i v_j}\) denotes the element in the identity matrix. We define the element matrix \(H\) by $h_{v_i v_j} = \big[ 1 - (1 - \gamma_{v_i v_j}) p^{A}_{v_i} \big] b_{v_i v_j}$,
and denote its largest eigenvalue as \(\lambda_{\max}(H)\).

Therefore,~\eqref{eq:11} reduces to the problem of finding the largest eigenvalue of the matrix, and the critical threshold \(\beta_c\) is given by:
\begin{equation}
\label{eq:12}  
\beta_c = \frac{\mu}{\lambda_{\max}(H)}.
\end{equation}

According to~\eqref{eq:11} and~\eqref{eq:12}, the epidemic outbreak threshold depends on the recovery rate \(\mu\), the differential reception intensity \(\gamma_{v_i v_j}\) of node \(v_i\) to information from each neighbor \(v_j\), and the fraction of nodes in the aware state \(p^A_{v_i}\). Furthermore, the network topology of the physical layer also affects the epidemic threshold.

\begin{figure}[!t]
\centering
\includegraphics[width=0.47\textwidth]{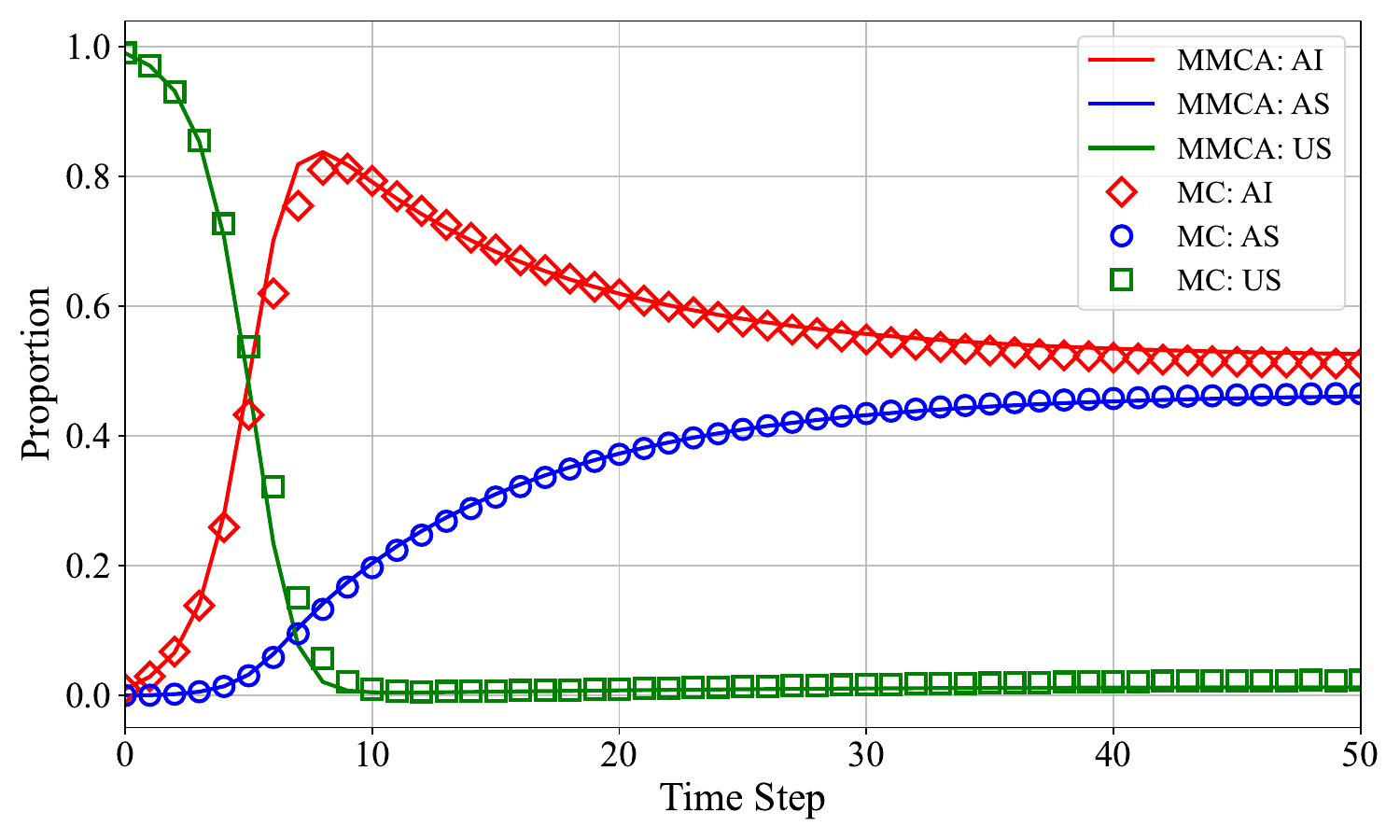}
\caption{Proportion changes over time. The red curve and the red hollow diamonds denote the proportion of AI-state nodes under the MMCA and MC methods, respectively. The blue curve and the blue hollow circles correspond to the proportion of AS-state nodes under the MMCA and MC methods, respectively. The green curve and the green hollow squares represent the proportion of US-state nodes under the MMCA and MC methods, respectively. In the physical layer, the parameters are set as follows: the initial infected node ratio is 1\%, disease transmission rate $\beta = 0.5$, and $\mu = 0.05$. In the cyber layer, the parameters are set as follows: $\lambda = 0.01$, $\lambda^* = 0.01$, $\delta = 0.05$, $\alpha = 1$, and $\eta = 1$.
}
\label{fig_3}
\end{figure}

\section{Simulations}
\label{sec4:simulations}
In this section, we perform extensive simulations to validate the theoretical framework and investigate the impact of information diffusion and interpersonal relationships on epidemic dynamics in the CPS. The simulations consist of four parts: (i) analysis of the temporal evolution of node state densities; 
(ii) comparison between theoretical predictions and simulation results; (iii) investigation of the effect of interpersonal relationships on epidemic dynamics; and (iv) examination of the model dynamics under various parameter combinations. All simulations are implemented in Python 3.12. For all the following simulations, in the cyber layer, we set the total number of nodes to $N=200$ and the number of hyperedges to $M=500$. We generate a connected mixed hypergraph using the method described in Sec.~II. The probability of connecting any pair of nodes within a hyperedge is set to $p_1=0.5$. In the physical layer, we adopt a WS small-world network model, generated with the 
\textit{watts\_strogatz\_graph()} function from the \textit{networkx} library. The initial infected nodes are randomly selected with the \textit{random.choice()} function from the 
\textit{NumPy} package. The total number of nodes is also 200, and $1\%$ of them are initially infected. Each node has $K=4$ neighbors, and the rewiring probability is $p_2=0.5$. Nodes in the physical layer correspond one-to-one with those in the cyber layer.

\subsection{Time Evolution of Node State Densities}
To investigate the dynamic evolution of node states in the coupled UAU--SIS spreading model and to evaluate the predictive accuracy of the MMCA, we analyze the time evolution of node state densities. Fig.~3 shows the temporal proportions of nodes in the three possible states—AI, AS, and US—obtained from both the MMCA and MC simulations.

\begin{figure}[!t]
\centering
\includegraphics[width=0.47\textwidth]{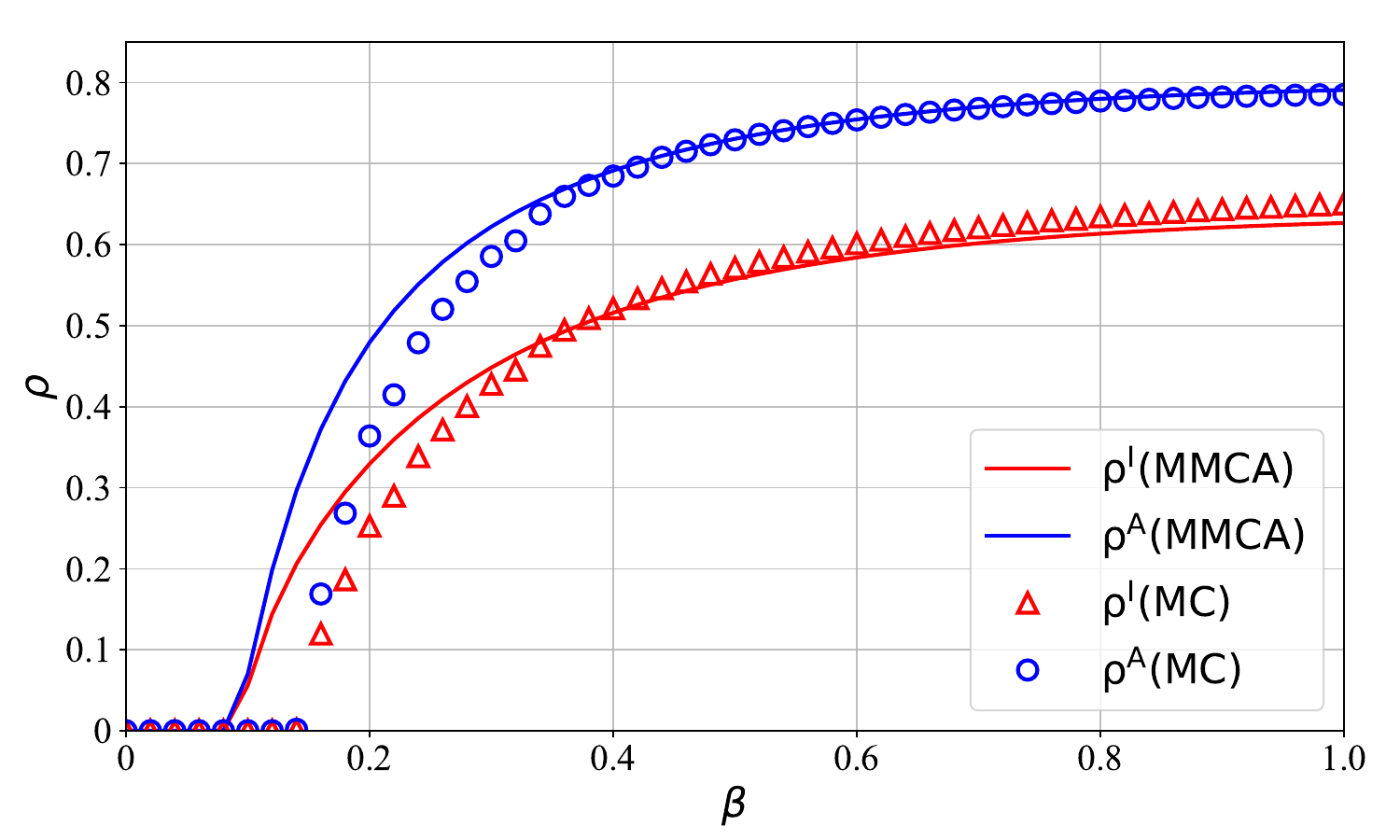}
\caption{Comparison of steady-state density between MC and MMCA. The red curve and red hollow triangles represent the infection densities obtained from the MMCA and MC methods, respectively, while the blue curve and blue hollow circles denote the awareness densities under the two approaches. Parameters for the physical layer are set as follows: the initial infected node ratio is 1\% and the recovery rate is $\mu = 0.4$. In the cyber layer, the parameters are set as follows: $\lambda = 0.1$, $\lambda^* = 0.1$, $\delta = 0.8$, $\alpha = 1$, and $\eta = 1$. The infection rate $\beta$ is varied from 0 to 1 with an increment of 0.02. All MC results are averaged 200 independent simulation runs.
}
\label{fig_4}
\end{figure}

As shown in Fig.~3, the system evolves into a stable state within a limited number of time steps. The densities of various node states stabilize after around 40 steps, indicating that the system has reached a steady-state phase. During this process, the proportion of nodes in the US state decreases sharply, reflecting the rapid spread of awareness to most nodes in the early stage under the influence of the information diffusion mechanism. The proportion of AI-state nodes rises steeply at first and then gradually declines, while the proportion of AS-state nodes increases steadily before eventually stabilizing. These dynamics demonstrate that awareness diffusion in the cyber layer plays a crucial role in suppressing disease transmission in the physical layer. Furthermore, the MMCA results exhibit strong agreement with MC simulations for the propagation process at a fixed transmission rate of $\beta = 0.5$. Both methods yield highly consistent trends and final densities across the three node states, indicating that the proposed theoretical framework can accurately capture the dynamics of a single propagation process while demonstrating strong predictive power and numerical stability.

\begin{figure}[!t]
\centering
\includegraphics[width=0.47\textwidth]{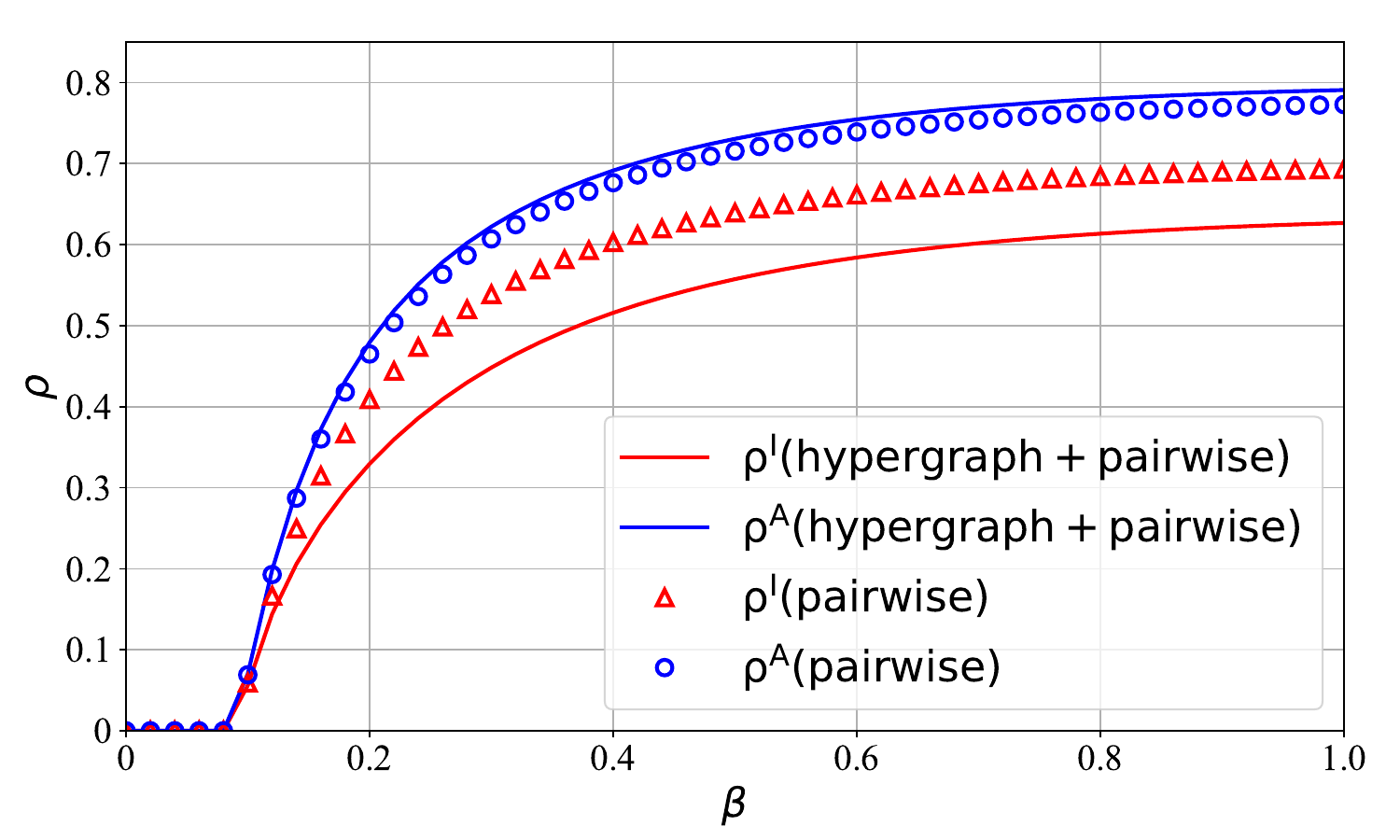}
\caption{Effect of higher-order interactions: comparison with a pairwise baseline. The red curve and red hollow triangles represent the infection densities obtained under the mixed hypergraph network and the pairwise network, respectively, while the blue curve and blue hollow circles denote the awareness densities under the two networks. The physical-layer parameters for both networks are set as follows: the initial infected node ratio is 1\% and the recovery rate is $\mu = 0.4$. In the cyber layer, the parameters are set as follows: $\lambda = 0.1$, $\lambda^* = 0.1$, $\delta = 0.8$, $\alpha = 1$, and $\eta = 1$. The infection rate $\beta$ is varied from 0 to 1 with an increment of 0.02.
}
\label{fig_5}
\end{figure}

In summary, we analyzed the temporal evolution of node-state proportions during the propagation process with a fixed infection rate $\beta = 0.5$. The system evolved into a stable state within a very finite number of time steps, and the MMCA results show a strong agreement with the MC simulations. Based on these results, we will further simulate the coupled dynamics in cyber-physical systems and investigate the epidemic outbreak threshold. In addition, we will analyze how the steady-state infection and awareness densities change under different transmission rates and assess the consistency between theoretical predictions and simulation results throughout the propagation process.

\begin{figure}[!t]
\centering
\includegraphics[width=0.47\textwidth]{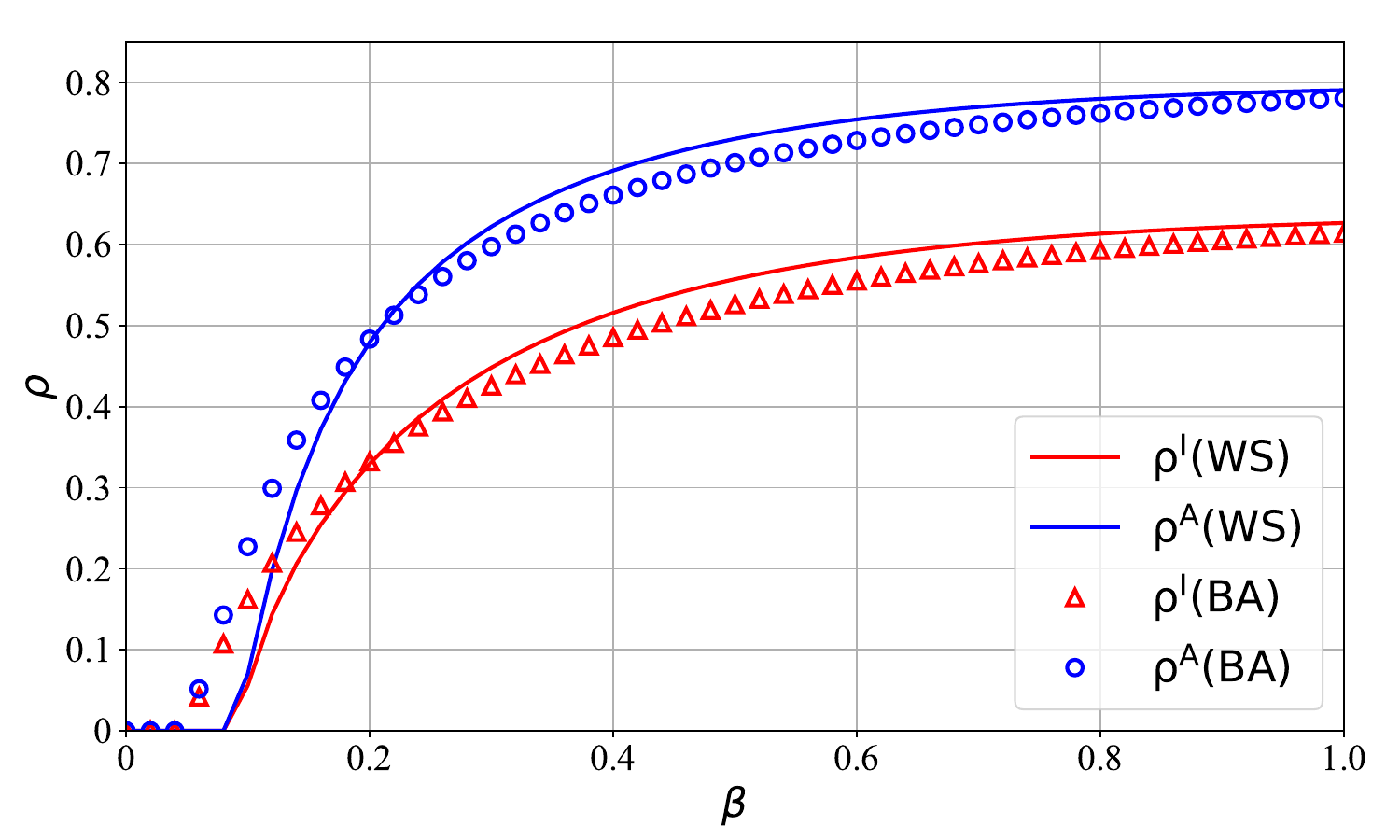}
\caption{Comparison between WS and BA Contact Networks. The red curve and red hollow triangles represent the infection densities obtained under the WS network and BA network, respectively, while the blue curve and blue hollow circles denote the awareness densities under the two networks. The physical-layer parameters for both networks are set as follows: the initial infected node ratio is 1\%, the recovery rate is $\mu = 0.4$, and the average degree is $k = 4$. In the cyber layer, the parameters are set as follows: $\lambda = 0.1$, $\lambda^* = 0.1$, $\delta = 0.8$, $\alpha = 1$, and $\eta = 1$. The infection rate $\beta$ is varied from 0 to 1 with an increment of 0.02.
}
\label{fig_6}
\end{figure}

\subsection{Comparison of Theoretical Predictions and Simulation Results}
To verify the effectiveness of the proposed model in simulating epidemic dynamics in the CPS, in this subsection, we employ the MC simulation method to investigate the epidemic spreading dynamics in the CPS and compare the outcomes with MMCA results. Specifically, we analyze the steady-state densities of infected and aware nodes for different infection rates $\beta$. The MMCA iterates serially by numerically solving state probability equations, whereas the MC simulations are parallelized due to the independence of random realizations. For both methods, the iterative process terminates once the difference in steady-state densities between successive iterations falls below a predefined convergence threshold.

Fig.~4 displays the results obtained from the MMCA and the MC simulations. As evident from Fig.~4, the two approaches exhibit a strong agreement in both infection and awareness densities. The analysis of the data shows that the average deviation between the two methods is approximately 2.9\%. This high consistency further corroborates the validity and accuracy of the proposed theoretical framework. Moreover, our results also indicate that with increasing infection rates, both the awareness density and the infection density increase simultaneously in a nonlinear manner, highlighting the nonlinear coupling dynamics between information diffusion and epidemic propagation in the UAU-SIS model. When the infection rate $\beta$ is relatively high, the results of the MC simulation and the MMCA method remain closely aligned. This suggests that in scenarios where the infection spreads extensively, MMCA can effectively approximate the dynamics captured by the MC simulations. However, when $\beta$ is low, a noticeable discrepancy emerges: the epidemic threshold estimated by the MC simulation is substantially higher than that predicted by MMCA. This divergence stems from the fundamental methodological differences. MC is a stochastic, event-driven approach that represents the system as a discrete process. At low infection rates, initially infected nodes in MC simulations may recover quickly, driving the system into an absorbing state—where no infected individuals remain and further transmission is impossible. In contrast, MMCA is a continuous, deterministic approximation that tracks the probability of each node being in the US, AS, or AI state. Even at extremely low infection probabilities, MMCA maintains non-zero values in a floating-point form, allowing weak transmission to persist and preventing the system from entering an absorbing state. This intrinsic difference explains why MMCA tends to underestimate the epidemic threshold compared to MC simulations in the low-$\beta$ regime.

In addition, it is important to note that when the infection rate falls below the epidemic threshold predicted by MMCA, the disease cannot spread. However, once the infection rate surpasses this threshold, an outbreak will occur. As shown in Fig.~4, under the same simulation conditions, the infection density and the awareness density simultaneously present an outbreak state, confirming the interaction and synergy between information diffusion and disease propagation in the CPS. Furthermore, both densities rise rapidly during the early stages but slow down when the system approaches saturation. This is because, at the outset, only a few nodes are infected or aware, while the majority remain susceptible. Infected or aware nodes thus have many unaffected neighbors, creating abundant opportunities for propagation and driving a rapid initial growth. As the process evolves, however, most nodes become infected or aware, leading to a rapid decline in the number of susceptible nodes. Consequently, new transmission paths become limited and the system gradually approaches either a stable state or an absorbing state, resulting in a slowdown in the overall growth rate of the infection and awareness densities.

The simulation results demonstrate that the MMCA can accurately predict the infection and awareness densities within the network, thereby confirming the effectiveness of the proposed model in simulating epidemic dynamics in the CPS. To further investigate the coupled dynamics between information diffusion and epidemic spreading, Section IV-E examines how interpersonal relationships influence the evolution of the epidemic, and how individuals adjust their infection risk by taking self-protective measures based on their relationships with neighboring nodes.

\begin{figure}
  \centering
  \subfloat[]{
    \includegraphics[width=0.47\textwidth]{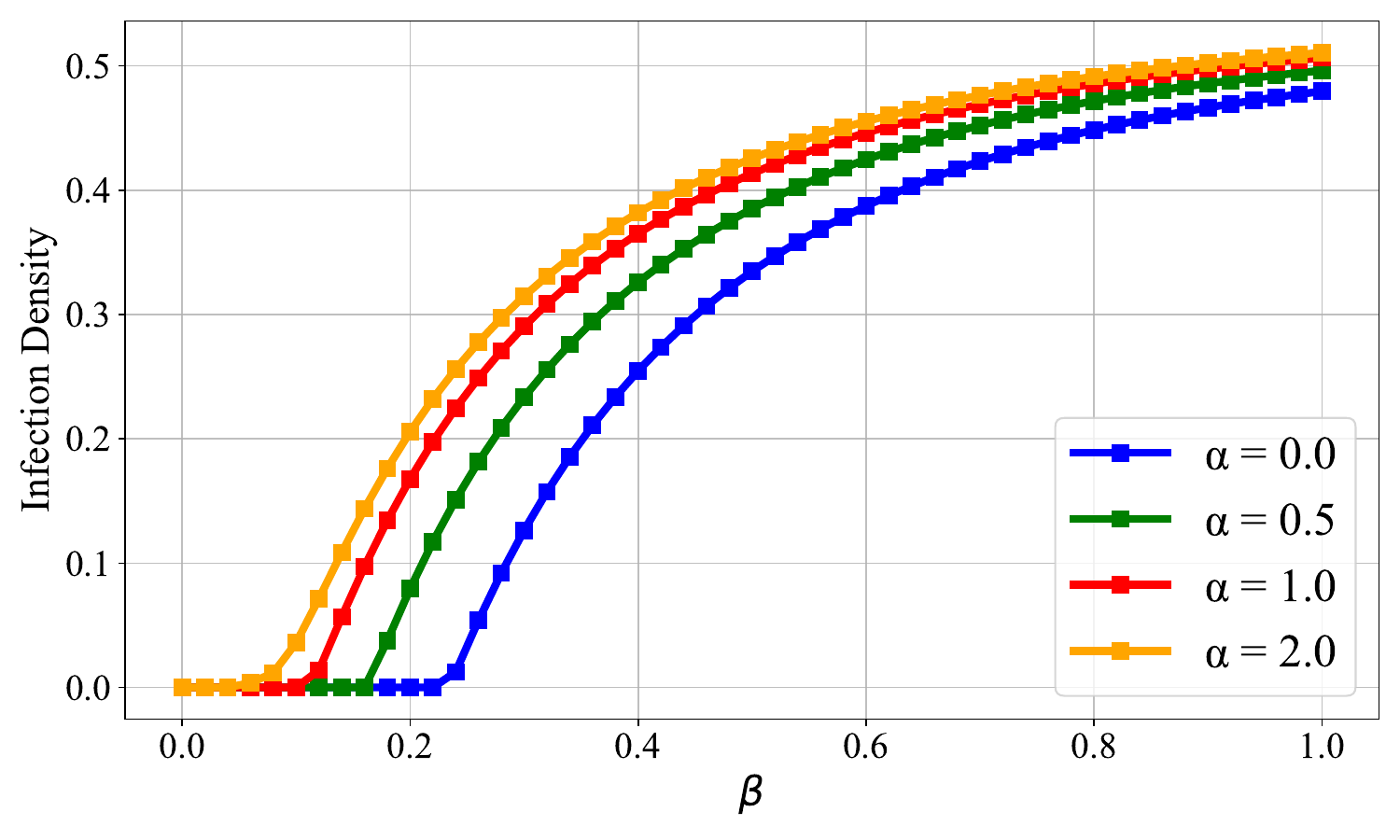}
    \label{fig:alpha_infection}
  }
  \hfill
  \subfloat[]{
    \includegraphics[width=0.47\textwidth]{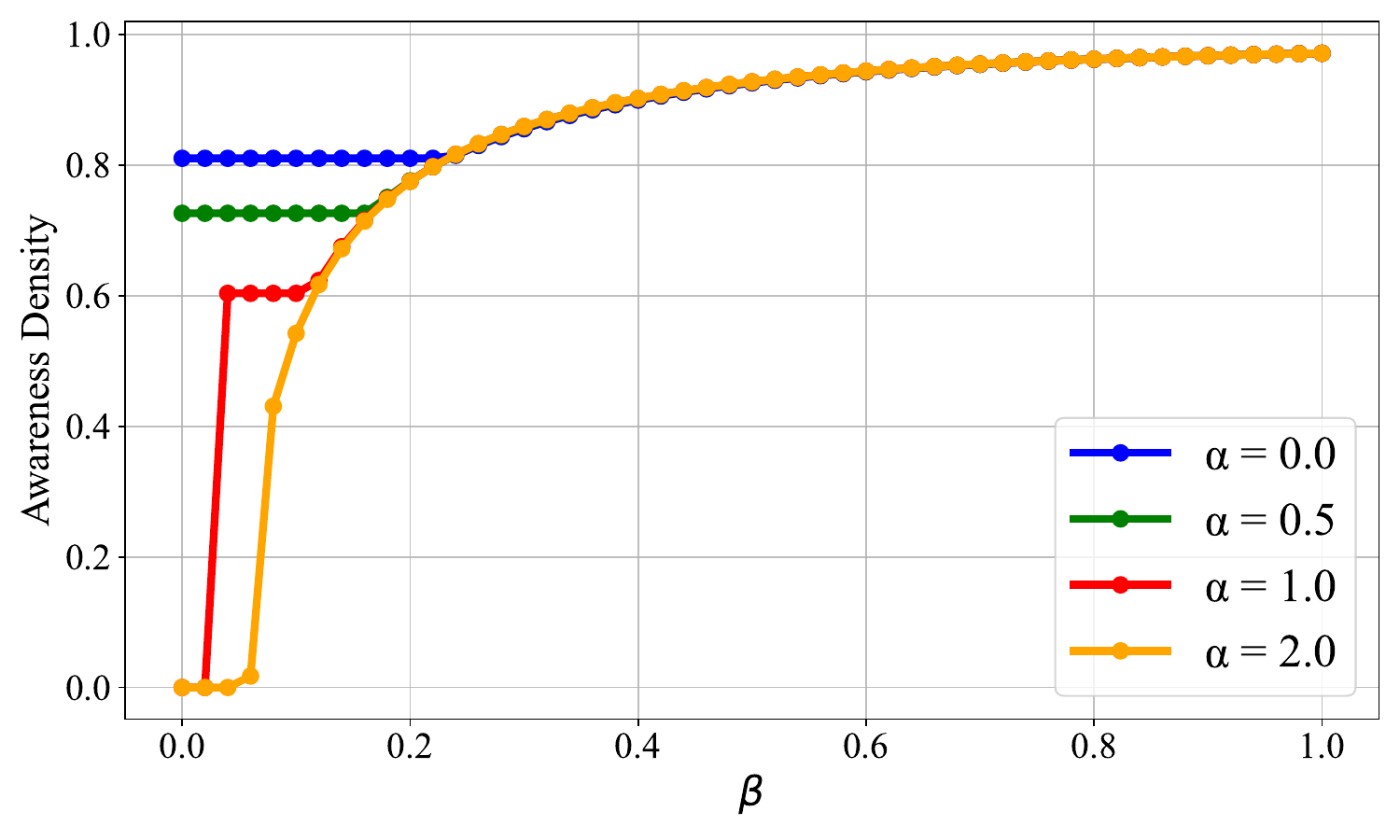}
    \label{fig:alpha_awareness}
  }
  \caption{Effect of an individual's relationship strength with neighbors on infection and awareness densities. The blue, green, red, and orange solid square lines represent infection densities for $\alpha = 0.0$, $0.5$, $1.0$, and $2.0$, respectively. Similarly, the blue, green, red, and orange solid circle lines correspond to the awareness densities for the same values of $\alpha$. Parameters for the physical layer are set as follows: the initial infected node ratio is 1\%, and the recovery rate is $\mu = 0.2$. Parameters for the cyber layer are set as follows: $\lambda = 0.3$, $\lambda^* = 0.1$, $\delta = 0.2$, and $\eta = 1$. The infection rate $\beta$ varies from 0 to 1 with an increment of 0.02. (a) Infection density changes with different values of $\alpha$. (b) Awareness density changes with different values of $\alpha$.}
  \label{fig_7}
\end{figure}

\begin{figure}
  \centering
  \subfloat[]{
    \includegraphics[width=0.47\textwidth]{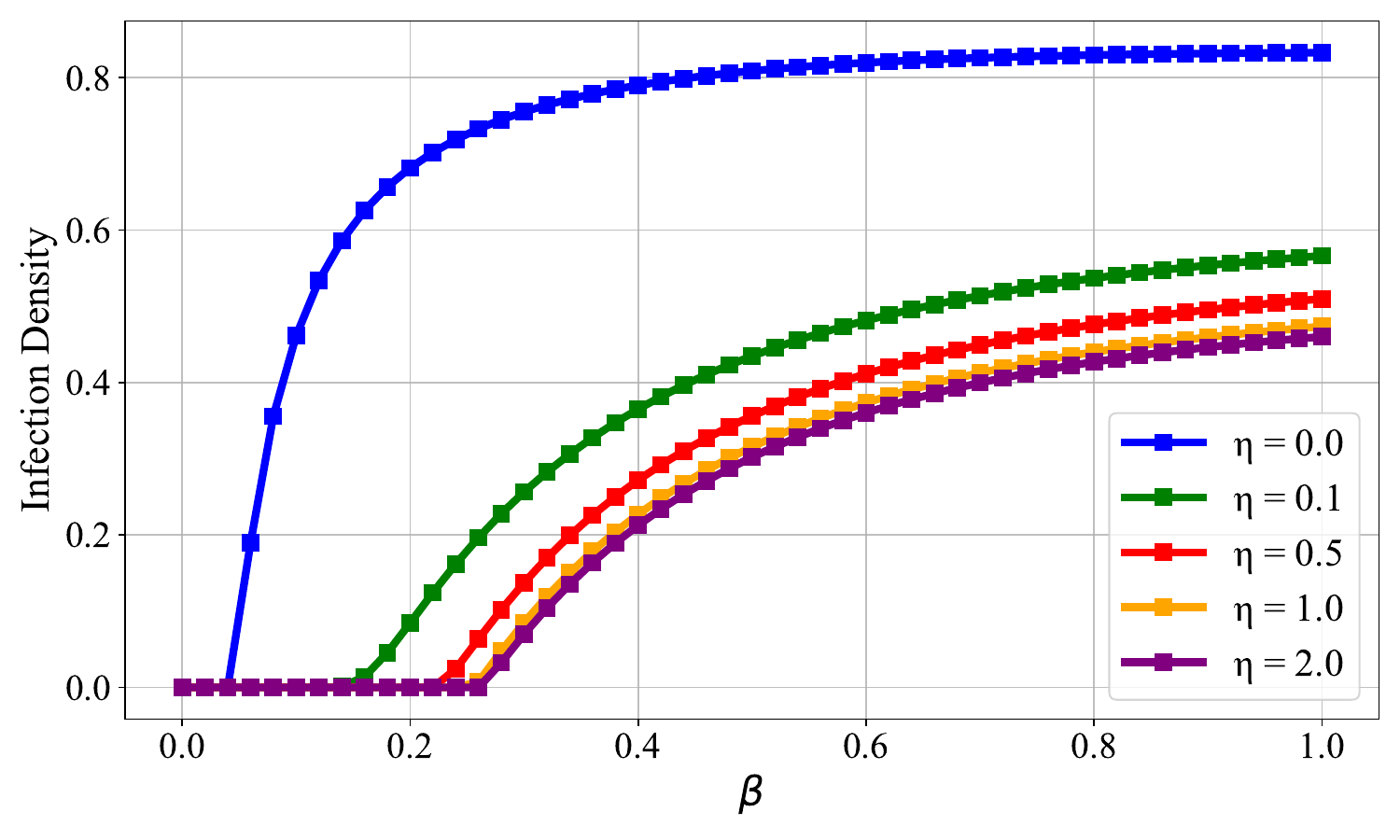}
    \label{fig:eta_infection}
  }
  \hfill
  \subfloat[]{
    \includegraphics[width=0.47\textwidth]{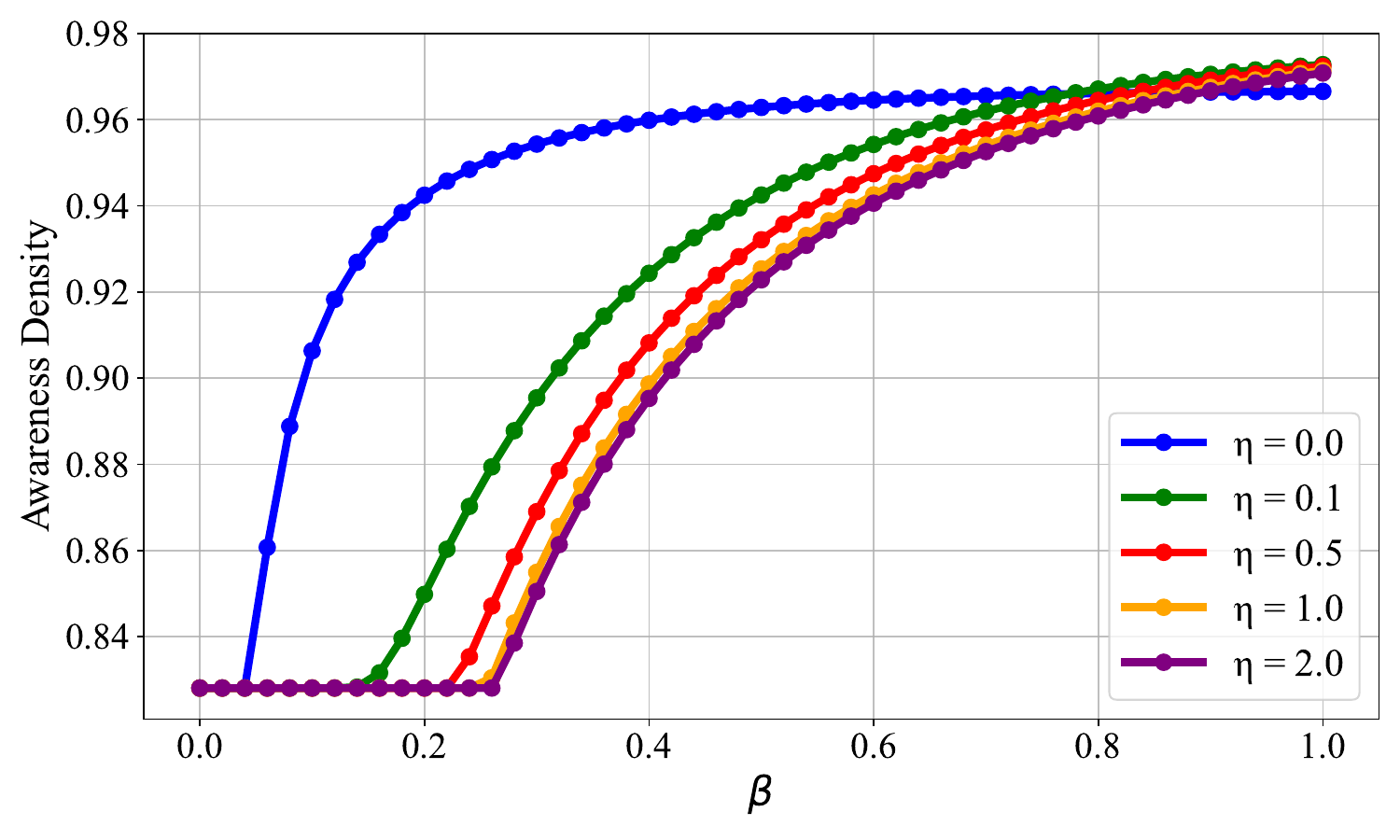}
    \label{fig:eta_awareness}
  }
  \caption{Impact of individuals' self-protection measures against different neighbors on infection and awareness densities. The blue, green, red, orange, and purple solid square lines represent the infection density for $\eta = 0.0$, $0.1$, $0.5$, $1.0$, and $2.0$, respectively. Similarly, the blue, green, red, orange, and purple solid circle lines represent the awareness densities for the same values of $\eta$. Parameters for the physical layer are set as follows: the initial infected node ratio is 1\%, and the recovery rate is $\mu = 0.2$. Parameters for the cyber layer are set as follows: $\lambda = 0.6$, $\lambda^* = 0.6$, $\delta = 0.2$, and $\alpha = 1$. The infection rate $\beta$ varies from 0 to 1 with an increment of 0.02. (a) Infection density changes with different values of $\eta$. (b) Awareness density changes with different values of $\eta$.}
  \label{fig_8}
\end{figure}

\subsection{Effect of Higher-Order Interactions: Comparison with a Pairwise Baseline}
To evaluate the specific contribution of higher-order interactions in the information layer, we compare the spreading behavior of the proposed mixed hypergraph-based model with that of a pairwise baseline model under identical conditions. In the baseline model, the pairwise awareness transmission mechanism and all model parameters remain the same as in the original model, while the higher-order interaction term is removed. Consequently, awareness diffusion is driven solely by pairwise contacts between individuals. Based on the MMCA framework, we analyze the steady-state densities of infected and aware nodes in the two models to quantify the impact of higher-order interactions on the coupled information--epidemic dynamics. This comparison allows us to isolate the role of higher-order interactions and assess the additional contribution introduced by the hypergraph formulation.

As shown in Fig.~5, when the infection transmission rate $\beta$ exceeds a certain threshold, an epidemic outbreak occurs and the densities of infected and aware nodes increase rapidly before gradually approaching steady-state values. It can be observed that the outbreak thresholds of the two models are nearly identical. This is because higher-order information transmission requires the simultaneous participation of multiple aware individuals, whereas pairwise transmission can occur through a single aware neighbor. In the early stage of spreading, the density of aware nodes is very low, making higher-order interactions difficult to activate. As a result, the spreading dynamics near the outbreak threshold are mainly dominated by pairwise transmission, leading to similar threshold values in the two models. However, once the spreading process develops and the density of aware nodes increases, higher-order interactions become more effective in promoting awareness diffusion, which further suppresses epidemic propagation. Compared with the pairwise baseline model, the mixed hypergraph-based model consistently produces a lower infection density and a higher awareness density over a wide range of transmission rates. This difference can be attributed to the higher-order interactions in the information layer, which enhance awareness diffusion through group interactions and thus strengthen the suppression effect on epidemic spreading. As $\beta$ further increases, although the steady-state densities in the two models gradually become closer, the hypergraph-based model still maintains a lower infection prevalence. These results demonstrate that higher-order interactions play an important role in enhancing awareness diffusion and reducing epidemic prevalence in the coupled information--epidemic dynamics.

\subsection{Comparison between WS and BA Contact Networks}
To examine the influence of network topology on the coupled information--epidemic dynamics, we compare the spreading behavior of the UAU--SIS model on two representative contact networks with distinct structural properties, namely the WS network and the BA network. The WS network is characterized by high clustering and a relatively homogeneous degree distribution, whereas the BA network exhibits strong degree heterogeneity due to the presence of hub nodes. Based on the MMCA framework, we analyze the dynamic behavior of the system on these two network structures to investigate the impact of network heterogeneity.

As shown in Fig.~6, when the infection transmission rate $\beta$ exceeds a certain threshold, an epidemic outbreak occurs, after which the densities of infected and aware nodes increase rapidly and gradually approach steady-state values. Compared with the WS network, the BA network exhibits a lower outbreak threshold and a faster increase in node densities, which can be attributed to its strong degree heterogeneity and the presence of hub nodes that facilitate the spreading process. However, as $\beta$ further increases, the steady-state densities obtained on the two network structures become very close, indicating that the overall spreading behavior of the coupled UAU--SIS model remains qualitatively consistent across different contact topologies. These results suggest that the proposed model is robust to variations in network heterogeneity.

\subsection{Impact of Perception Mechanisms on Epidemic Dynamics with Interpersonal Relationships}
In this subsection, we investigate the impact of interpersonal relationships and the varying self-protection strategies employed by individuals in response to neighbors with different relational strengths on the spread of epidemics. The infection and awareness densities under different interpersonal relationships and protection strategies are illustrated in Figs.~7 and~8. 

\begin{figure*}
  \centering
  \subfloat[]{
    \includegraphics[width=0.47\textwidth]{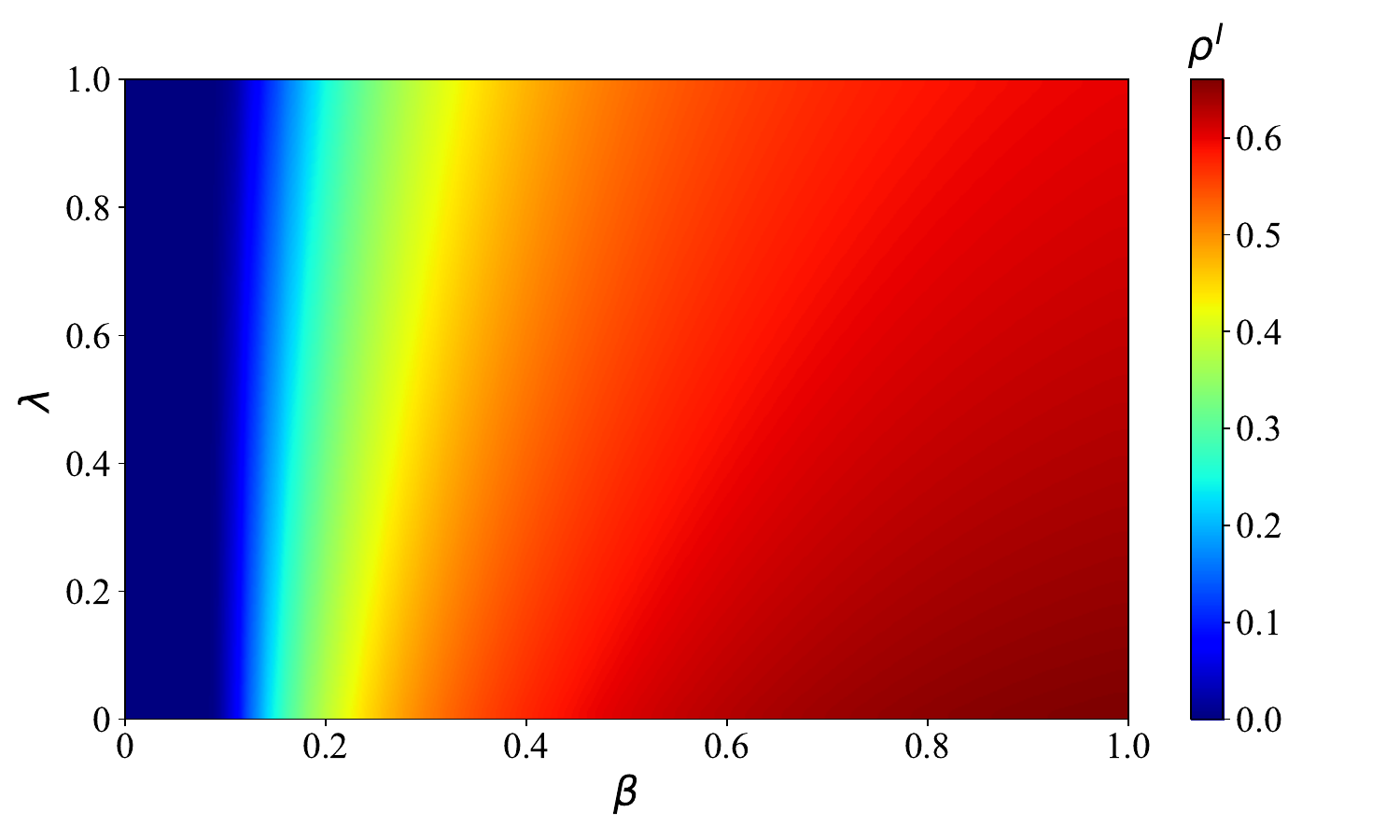}
    \label{fig:lambda_infection}
  }
  \hfill
  \subfloat[]{
    \includegraphics[width=0.47\textwidth]{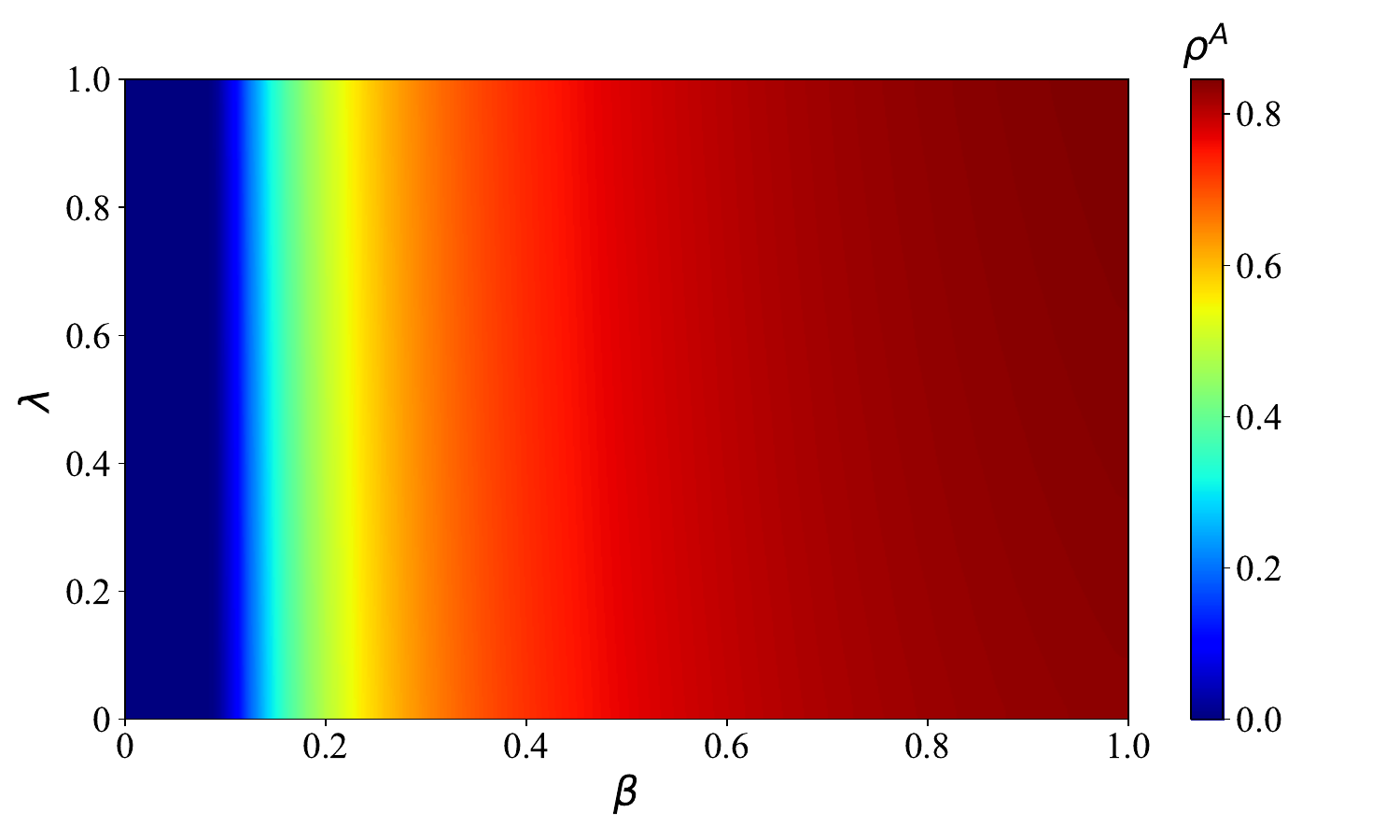}
    \label{fig:lambda_awareness}
  }
  \par\vspace{0.05em} 
  \subfloat[]{
    \includegraphics[width=0.47\textwidth]{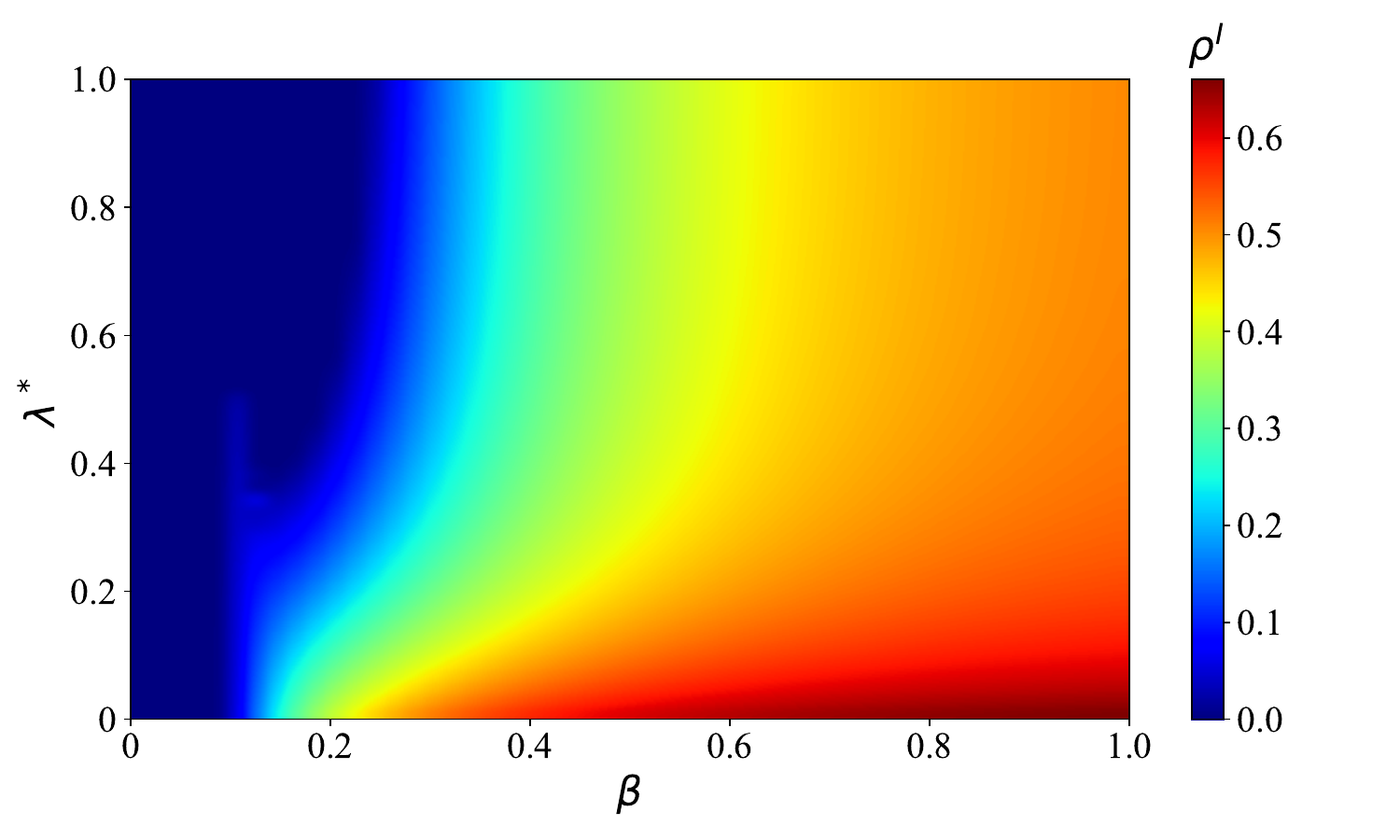}
    \label{fig:lambda^*_infection}
  }
  \hfill
  \subfloat[]{
    \includegraphics[width=0.47\textwidth]{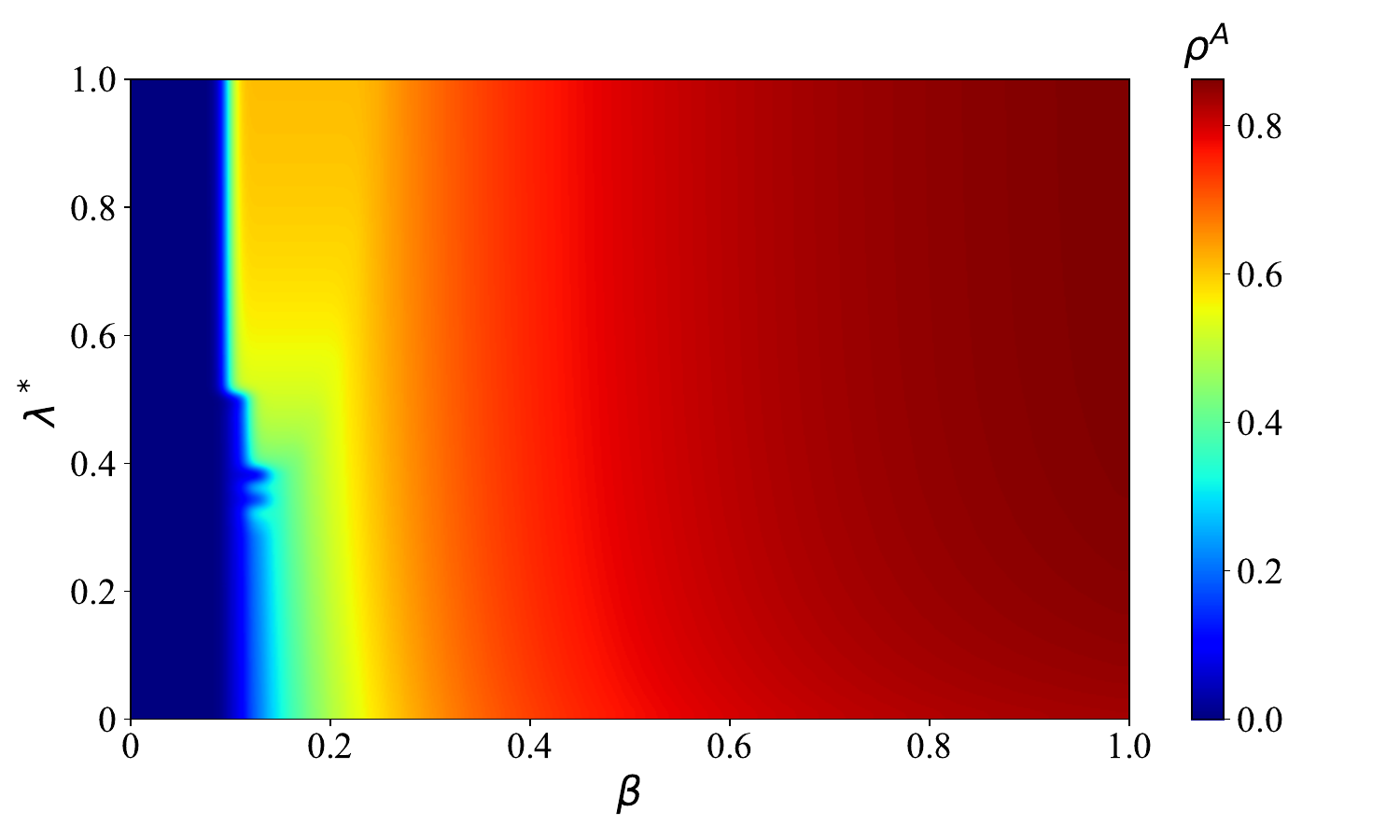}
    \label{fig:lambda^*_awareness}
  }
  \caption{Heatmaps of infection and awareness densities under different information propagation methods. (a) and (b) Infection and awareness densities, respectively, when only pairwise interaction information is considered. (c) and (d) Infection and awareness densities, respectively, when only higher-order interaction information is considered. In (a) and (b), physical-layer parameters are set as follows: the initial infection node ratio is 1\%, and the recovery rate is $\mu = 0.4$. The cyber layer parameters are: $\delta = 0.6$, $\lambda^*$ is not considered, $\alpha = 1$, and $\eta = 1$. In (c) and (d), physical-layer parameters are the same as above. The cyber-layer parameters are: $\delta = 0.6$, $\lambda$ is not considered, $\alpha = 1$, and $\eta = 1$. The color in each panel represents the corresponding density values distributed in a $50 \times 50$ grid.}
  \label{fig_9}
\end{figure*}

\begin{figure*}
  \centering
  \subfloat[]{
    \includegraphics[width=0.47\textwidth]{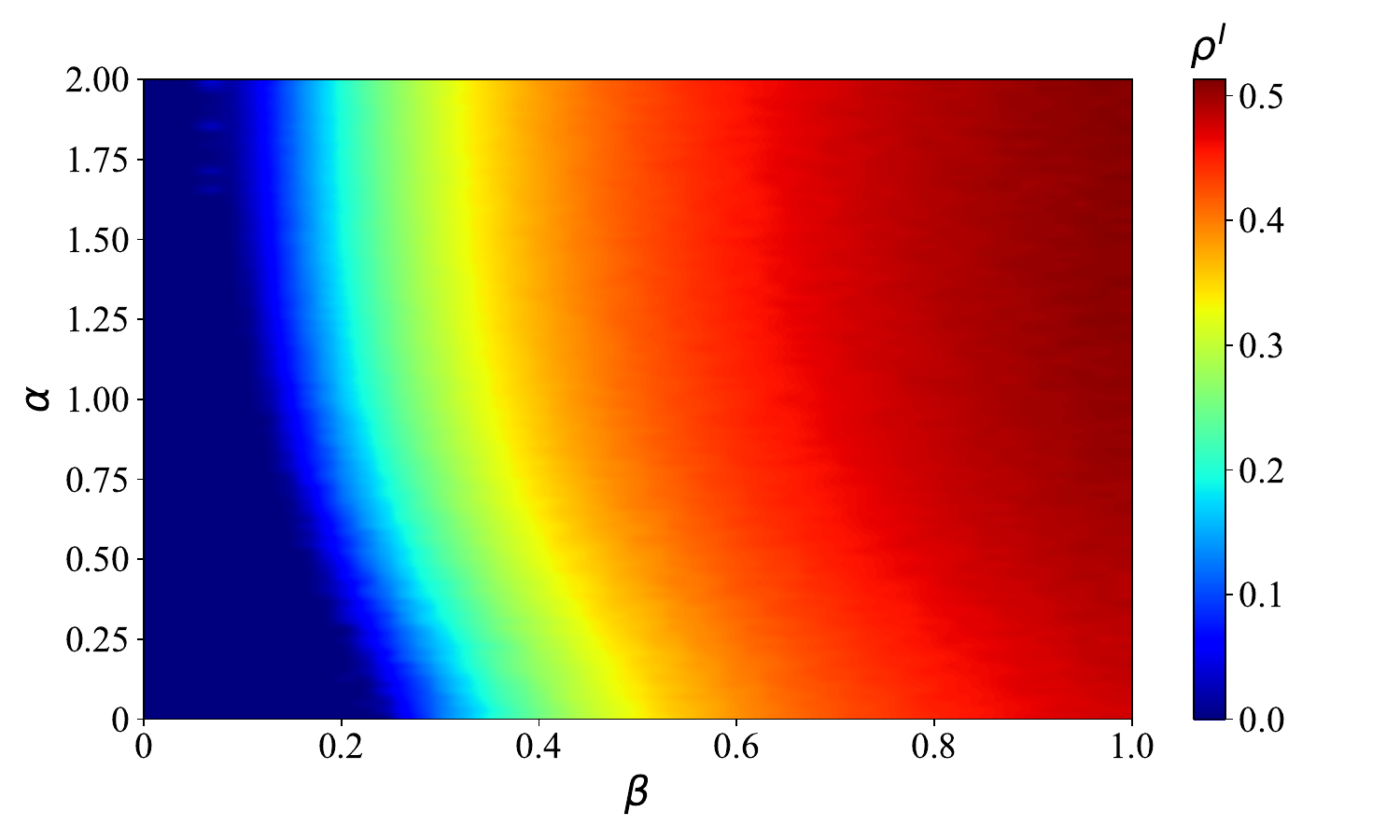}
    \label{fig:alpha_infection_heat map}
  }
  \hfill
  \subfloat[]{
    \includegraphics[width=0.47\textwidth]{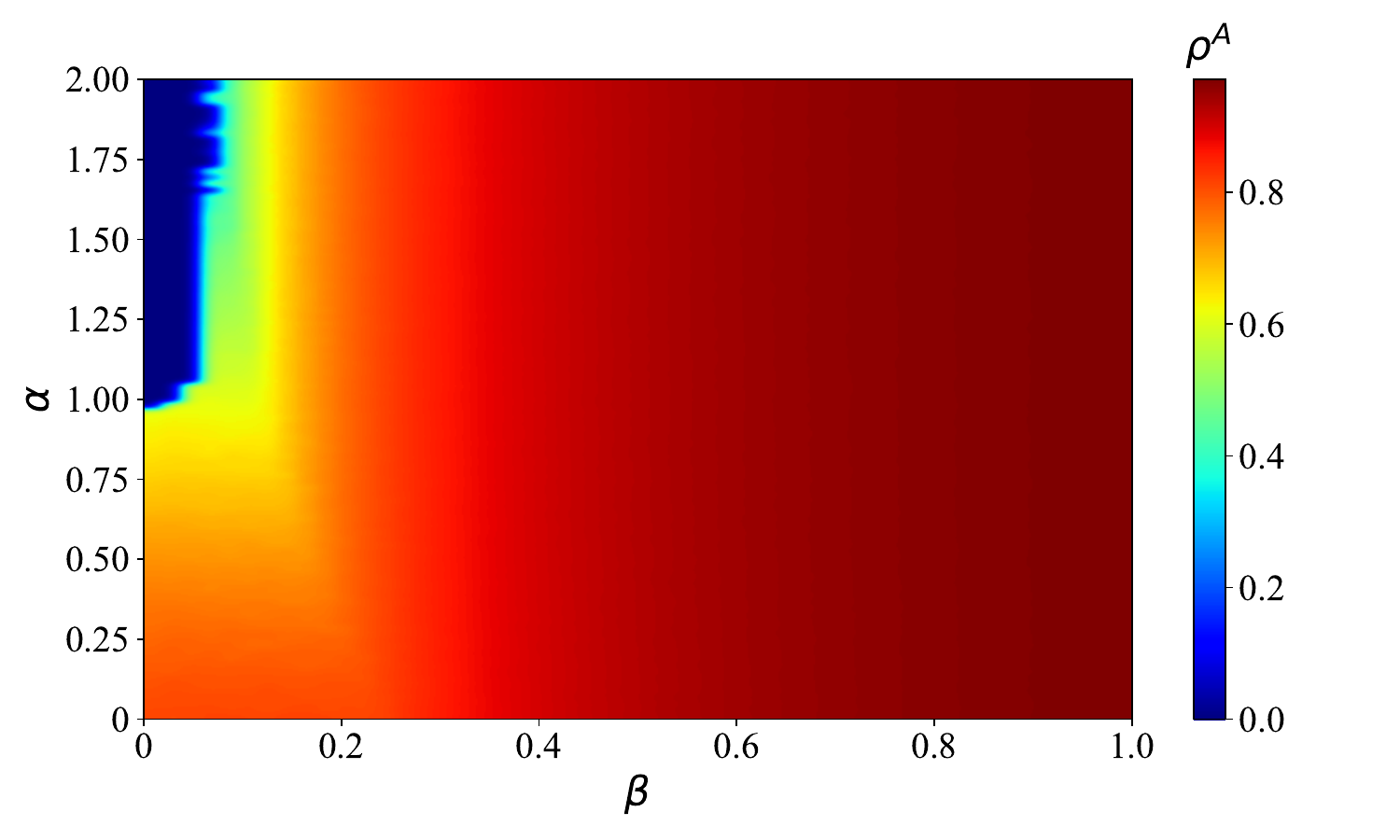}
    \label{fig:alpha_awareness_heat map}
  }
   \par\vspace{0.5em} 
  \subfloat[]{
    \includegraphics[width=0.47\textwidth]{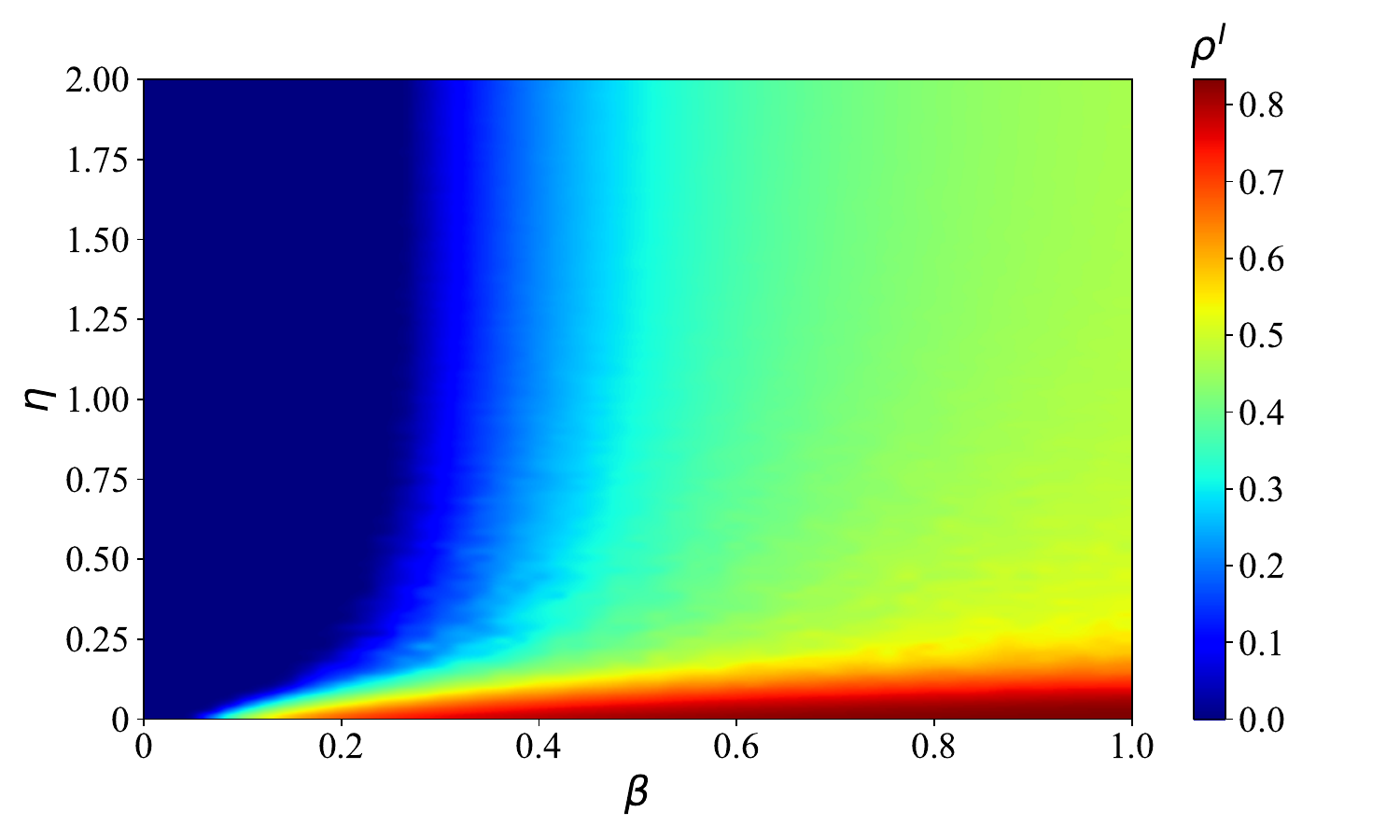}
    \label{fig:eta_infection_heat map}
  }
  \hfill
  \subfloat[]{
    \includegraphics[width=0.47\textwidth]{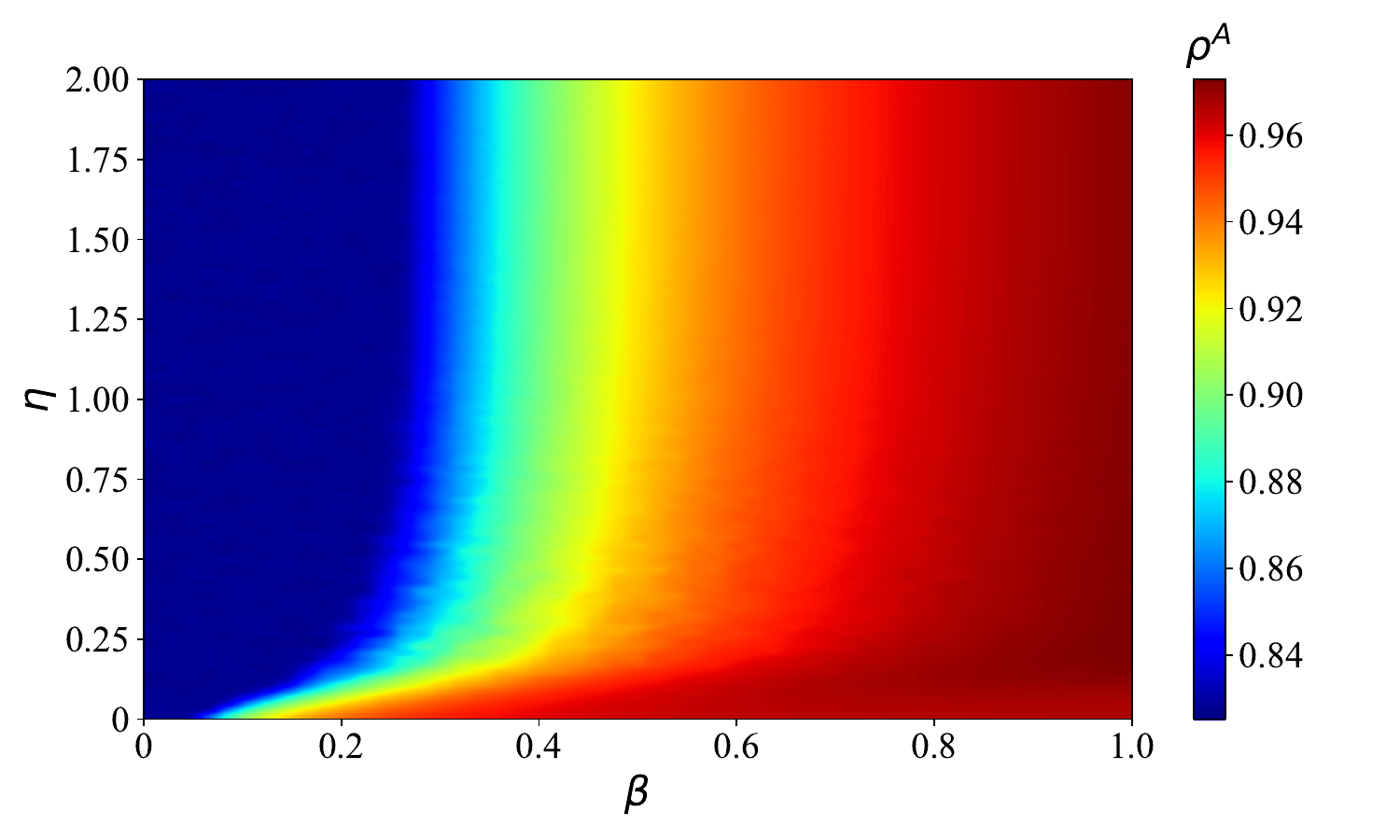}
    \label{fig:eta_awareness_heat map}
  }
  \caption{Heatmaps of infection and awareness densities under varying interpersonal relationship strengths and self-protection levels. (a) and (b) Infection and awareness densities, respectively, with the individual self-protection parameter fixed at $\eta = 1$. (c) and (d) Infection and awareness densities, respectively, with the interpersonal relationship strength parameter fixed at $\alpha = 1$. For (a) and (b), physical-layer parameters are set as follows: the initial infection ratio is 1\%, and the recovery rate is $\mu = 0.2$. The cyber-layer parameters are $\delta = 0.2$, $\lambda = 0.3$, and $\lambda^* = 0.1$. In (c) and (d), physical-layer parameters remain the same, while the cyber-layer parameters are set to $\delta = 0.2$, $\lambda = 0.6$, and $\lambda^* = 0.6$. The color in each panel represents the corresponding density values distributed in a $100 \times 50$ grid.}
  \label{fig_10}
\end{figure*}

In Fig.~7, we present the infection and awareness density curves as functions of the infection rate~$\beta$, for different values of the interpersonal relationship regulation factor~$\alpha$ (0.0, 0.5, 1.0, and 2.0), which governs the strength of the relationship between individuals and their neighbors. As shown in Figs.~7(a) and~(b), for the same $\beta$, increasing $\alpha$ leads to a higher infection density and a lower awareness density. This occurs because an increase in $\alpha$ weakens the strength of relationships between nodes, leading to a deterioration in the connections between individuals. Consequently, the information acceptance rate declines, reducing the proportion of awareness nodes. As a result, many individuals fail to take preventive measures in time and ultimately become infected. Moreover, as $\alpha$ increases, the infection threshold decreases, i.e., the epidemic will be more likely to break out. In addition, as shown in Fig.~7(b), when $\alpha = 0.0$ and $\alpha = 0.5$, the initial awareness density remains relatively high even before reaching the epidemic threshold. This is because at lower values of $\alpha$, individuals generally maintain strong connections with their neighbors, facilitating the widespread dissemination of information. When $\alpha = 1.0$, the weakening of interpersonal relationships is moderate, and the resulting relationship strength remains sufficient to sustain information diffusion within a certain range. As the number of aware nodes increases, higher-order information transmission in the hypergraph becomes increasingly effective. Once the awareness density exceeds a critical threshold, the awareness diffusion process is abruptly activated, driving the system from a ``non-aware'' state to a ``widely aware'' state. This abrupt change indicates a first-order phase transition. Such a feedback mechanism amplifies awareness once a critical mass of aware individuals is formed, leading to a discontinuous transition in awareness density. The emergence of this discontinuous transition can be attributed to the nonlinear feedback between awareness diffusion in the cyber layer and epidemic spreading in the physical layer. In particular, higher-order information transmission in the hypergraph requires the simultaneous presence of multiple aware individuals within a hyperedge, making awareness diffusion difficult to initiate when the awareness density is low. Once a critical mass of aware nodes is formed, this group-based transmission mechanism rapidly amplifies awareness, leading to a discontinuous transition in awareness density. In contrast, the theoretical analysis used to derive the epidemic threshold in Sec.~\ref{subsec:threshold} is based on a linear stability analysis near the disease-free equilibrium, which corresponds to the prediction of a continuous second-order phase transition. When $\alpha = 2.0$, the weakening of interpersonal relationships significantly suppresses information diffusion. As a result, the system remains in a ``non-aware'' state before reaching the epidemic threshold. Once $\beta$ exceeds the threshold, the epidemic starts to spread, and both the infection and awareness densities increase gradually.

In addition, Fig.~8 shows the infection and awareness density curves as functions of $\beta$ under different values of the self-protection intensity regulation factor~$\eta$ (0.0, 0.1, 0.5, 1.0, and 2.0), which quantifies the extent to which individuals adopt self-protection measures based on their neighbors' relationship strengths. As shown in Figs.~8(a) and~(b), for a given $\beta$, both the infection and awareness densities decrease as $\eta$ increases. A higher $\eta$ indicates that individuals adopt stronger self-protection measures when interacting with neighbors of varying relationship strengths, thereby reducing their likelihood of infection and resulting in a lower infection density. In the proposed model, it is assumed that once individuals become infected, they immediately acquire epidemic-related information. Consequently, the decrease in infection density directly leads to a reduction in awareness density. Furthermore, as $\eta$ increases, the epidemic outbreak threshold rises, i.e., it is more difficult for an epidemic to break out.

In summary, we have examined the effects of interpersonal relationships and the extent to which individuals adopt self-protective measures in response to neighbors with varying relationship strengths on the spread of epidemics. In Section IV-F, we will analyze the combined effects of the following factors: pairwise interactions and the infection rate~$\beta$, higher-order interactions and~$\beta$, the strength of interpersonal relationships and~$\beta$, and the extent of self-protective behavior and~$\beta$.

\subsection{Combined Effects of Various Factors on Epidemic Spread}
In this subsection, we use heat maps to analyze the combined effects of the infection rate~$\beta$ and four key factors---pairwise interactions, higher-order interactions, interpersonal relationship strength, and the extent of individual self-protection---on epidemic spread. Figs.~9 and~10 show the corresponding results.

In Fig.~9, we present heatmaps showing infection and awareness densities as functions of $\lambda$ and $\lambda^*$ for different values of the infection rate $\beta$. In Fig. 9(a) and (b), the influence of higher-order interaction propagation is excluded, while in Fig. 9(c) and (d), the effect of pairwise interaction propagation is excluded.

As shown in Fig.~9(a), when $\beta$ falls below the epidemic threshold (approximately 0.1), the disease fails to spread. Once $\beta$ surpasses this threshold, the infection density increases with rising $\beta$ and decreasing $\lambda$. A similar trend is observed in Fig.~9(c), indicating that a higher $\beta$ promotes disease transmission, while higher values of $\lambda$ and $\lambda^*$ effectively suppress the epidemic spread. Comparing Figs.~9(a) and~(c), we find that, for a fixed $\beta$, variations in $\lambda^*$ lead to more significant changes in the steady-state infection density than variations in $\lambda$. Specifically, the average steady-state infection densities in Figs.~9(a) and~(c) are approximately 0.463 and 0.332, respectively, which supports our observation. Fig.~9(b) shows that when $\beta$ falls below the epidemic outbreak threshold, awareness density remains at a low level, indicating that information has not yet spread widely through the network. Once $\beta$ exceeds the threshold, the epidemic outbreak provides a source of information for the cyber layer, triggering rapid awareness diffusion. As $\beta$ increases, the awareness density rises sharply, tending to saturate around $\beta \approx 0.3$, and then remains stable. This occurs because, once the epidemic spreads broadly, infected individuals automatically acquire disease-related information, keeping awareness density high. Fig.~9(d) exhibits a similar trend. By comparing Figs.~9(b) and~(d), we observe that a higher infection rate $\beta$ leads to a higher final awareness density. Additionally, increasing the pairwise information transmission rate $\lambda$ and the higher-order transmission rate $\lambda^*$ further boosts awareness density. This trend is more obvious in Fig.~9(d). The average awareness densities in Figs.~9(b) and~(d) are approximately 0.643 and 0.673, respectively, indicating that both pairwise and higher-order transmission mechanisms can effectively promote the diffusion of disease-related information.

In Fig.~10, we present heatmaps illustrating infection and awareness densities as functions of the regulation factors $\alpha$ and $\eta$ for different $\beta$. Specifically, in Fig.~10(a) and (b), the results are obtained with $\eta$ fixed at 1, while in Fig.~10(c) and (d), the results are obtained with $\alpha$ fixed at 1.

As shown in Fig.~10(a), when $\beta$ falls below the epidemic threshold, the disease fails to spread. Once $\beta$ exceeds this threshold, the infection density increases with both $\beta$ and $\alpha$, indicating that higher infection rates and larger $\alpha$ values facilitate the spread of the disease. Specifically, for a fixed $\beta$, the steady-state infection density increases as $\alpha$ rises, while the steady-state awareness density decreases. This occurs because an increase in $\alpha$ weakens the connection strength between nodes, which deteriorates interpersonal relationships and reduces information acceptance. As a result, the proportion of aware individuals decreases. This phenomenon is also evident in Fig.~10(b), where many individuals fail to take timely protective measures and eventually become infected. Notably, in the upper-left region of the heatmap in Fig.~10(b), the awareness density drops to zero. Although the initial awareness density in the network is high, when $\alpha$ exceeds 1, the relationship strength between nodes is significantly reduced. This leads to a rapid breakdown of interpersonal connections and a severe decline in the ability to spread awareness information, ultimately resulting in a complete loss of awareness in that region. Fig.~10(c) shows that infection density increases with $\beta$ and decreases with $\eta$. This indicates that higher infection rates promote epidemic spread, while stronger individual self-protection behaviors (that is, higher $\eta$ values) effectively suppress it. Specifically, for a fixed $\beta$, both steady-state infection and awareness densities decrease as $\eta$ increases. This occurs because a higher $\eta$ indicates that individuals adopt stronger protective behaviors, thereby reducing the likelihood of infection and lowering the overall infection density. In the proposed model, individuals are assumed to immediately become aware of the epidemic upon infection. Thus, a decrease in infection density directly leads to a reduction in awareness density, as shown in Fig.~10(d).

Overall, stronger interpersonal relationships and proactive self-protection behaviors can effectively mitigate epidemic spread. Therefore, it is crucial in real life to maintain close contact with family and friends, stay informed about epidemic-related information through the media, and adopt effective self-protection measures. These actions not only maximize personal safety but also help safeguard the health of families, society, even the nation.

\section{Conclusion and Outlook}
\label{sec5:conclusion and outlook}
In summary, this paper presents a hypergraph structure for CPS that serves as a bidirectionally coupled network model for information diffusion and epidemic spreading. To capture relationship heterogeneity, we propose an adaptive perceptual protection mechanism based on Jaccard similarity. The study investigates the heterogeneous effects of interpersonal relationships on information awareness and disease transmission, and how individuals adjust their infection risk by adopting self-protection strategies based on neighborhood connections. Our study demonstrates that interpersonal relationships play a crucial role in epidemic spreading. The strength of these relationships influences individuals' receptiveness to epidemic-related information and determines whether they can promptly perceive this information and adopt self-protective behaviors. Due to the heterogeneity in relationship strength, individuals exhibit varying degrees of protective behavior, which in turn alters the overall dynamics of disease transmission. Specifically, closer interpersonal relationships lead to a higher proportion of aware individuals and a lower infection density. Moreover, more effective self-protection measures adopted by individuals contribute to lowering both infection and awareness densities. Heterogeneity in relationship strength also affects the epidemic threshold: closer interpersonal relationships raise this threshold, thereby making epidemic outbreaks less likely. Similarly, stronger individual protective behaviors also increase the outbreak threshold, further hindering the spread of the disease. Our results provide a theoretical basis for designing cognition-driven prevention and control strategies and advance the understanding of epidemic spreading dynamics within hypergraph structures.

However, several aspects of the model warrant further improvement. In real-world scenarios, both interpersonal and physical contact networks are inherently dynamic. New nodes may join the network~\cite{10521680}, i.e., individuals entering social groups or newly infected persons, while some existing nodes may leave due to factors like death, migration, or prolonged inactivity~\cite{10844908, zeng2025bursty}. To capture these dynamics, queuing theory could be used to model node birth and death processes within the network. Moreover, in this paper, we use a hypergraph to model higher-order information interactions to ensure the model's universality. Nevertheless, our hypergraph model can also be extended to higher-dimensional scenarios. Finally, the current protection mechanism is primarily driven by structural overlap among individuals. In real-world scenarios, protective behaviors may also be influenced by factors such as information credibility, trust in information sources, and individual resistance to certain information channels. Therefore, more complex behavioral mechanisms, including the active spread of misinformation and heterogeneous resistance to information reception, should be considered. Future models may introduce additional node states in the cyber layer, such as misinformed and apathetic nodes, to capture these behavioral and cognitive factors. These extensions will be explored in future work to further enhance the realism and applicability of the proposed framework.


%





\ifCLASSOPTIONcaptionsoff
  \newpage
\fi

\bibliographystyle{IEEEtran}
\bibliography{references}

\begin{IEEEbiography}[{\includegraphics[width=1in,height=1.25in,clip,keepaspectratio]{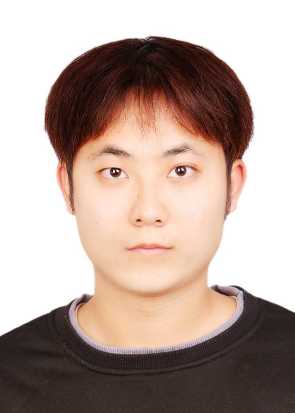}}]{Shanchao Peng}
received the B.E. degree from the College of Artificial Intelligence, Henan Agricultural University, Zhengzhou, China, in 2023. He is currently pursuing the M.S. degree with the College of Artificial Intelligence, Southwest University, Chongqing, China.

His research interests include complex networks, mathematical epidemiology, stochastic processes, and nonlinear science.
\end{IEEEbiography}

\begin{IEEEbiography}[{\includegraphics[width=1in,height=1.25in,clip,keepaspectratio]{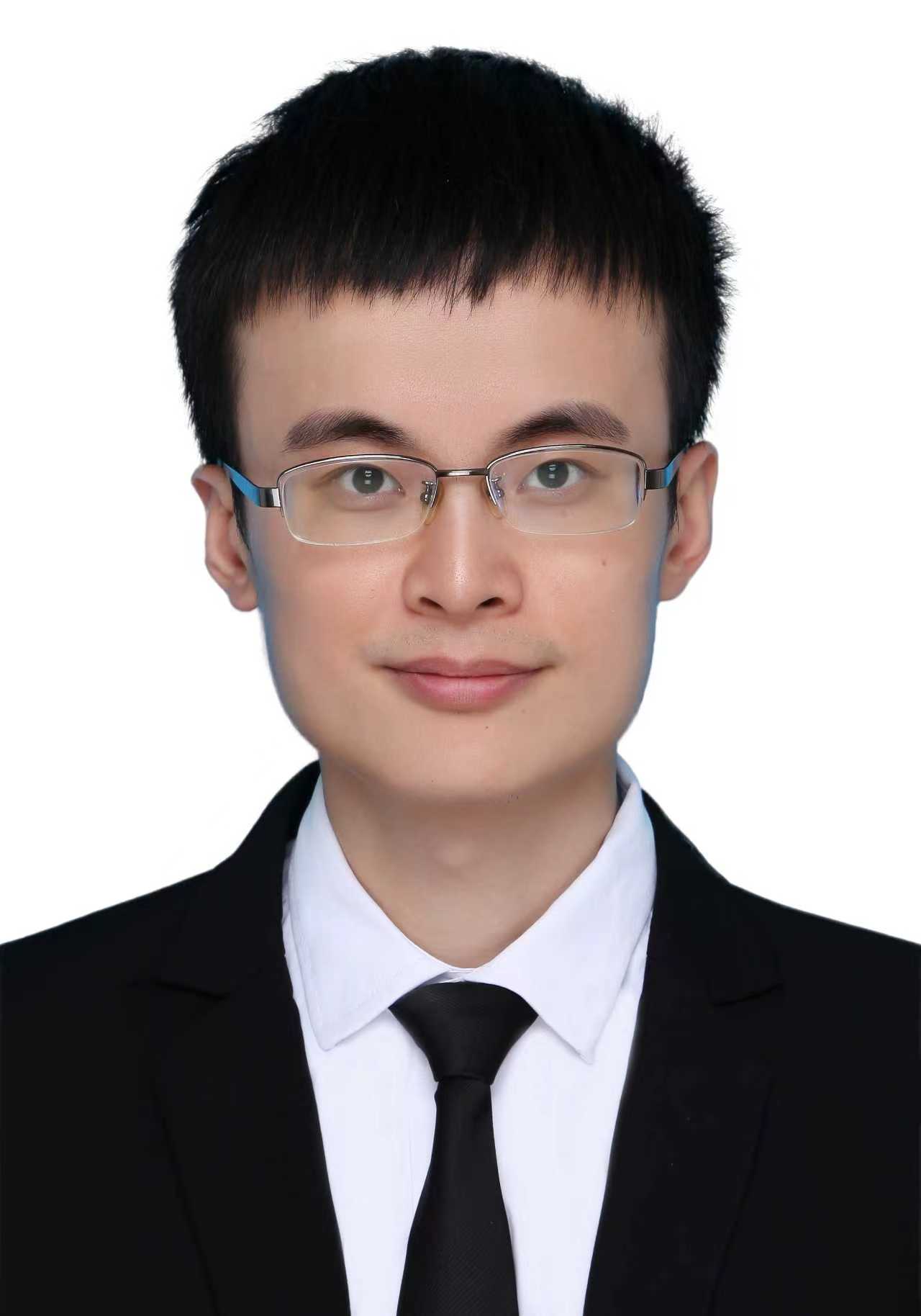}}]{Minyu Feng}
(Senior Member, IEEE) received his Ph.D. in Computer Science in 2018 through a joint program between the University of Electronic Science and Technology of China, Chengdu, China, and Humboldt University of Berlin, Berlin, Germany. Since 2019, he has been an Associate Professor at the College of Artificial Intelligence, Southwest University, Chongqing, China.

Dr. Feng has published more than 80 peer-reviewed papers in authoritative journals, including IEEE Transactions on Pattern Analysis and Machine Intelligence, IEEE Transactions on Systems, Man, and Cybernetics: Systems, and IEEE Transactions on Cybernetics. He is a Senior Member of the China Computer Federation (CCF) and the Chinese Association of Automation (CAA). Currently, he serves as a Subject Editor for Applied Mathematical Modelling, an Academic Editor for PLOS Computational Biology, an Editorial Advisory Board Member for Chaos, and an Editorial Board Member for Humanities \& Social Sciences Communications, Scientific Reports, and Discover Computing, among others. He also serves as a reviewer for Mathematical Reviews published by the American Mathematical Society.

Dr. Feng's research interests include complex systems, evolutionary game theory, computational social science, and mathematical epidemiology.
\end{IEEEbiography}

\begin{IEEEbiography}[{\includegraphics[width=1in,height=1.25in,clip,keepaspectratio]{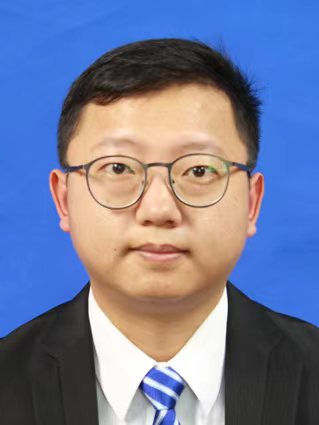}}]{Liang-Jian Deng}
(Senior Member, IEEE) received the B.S. and Ph.D. degrees in applied mathematics from the School of Mathematical Sciences, University of Electronic Science and Technology of China (UESTC), Chengdu, China, in 2010 and 2016, respectively.

He is currently a Research Fellow with the School of Mathematical Sciences, UESTC. From 2013 to 2014, he was a Joint-Training Ph.D. Student with Case Western Reserve University, Cleveland, OH, USA. In 2017, he was a Post--Doctoral Researcher with Hong Kong Baptist University (HKBU). In addition, he stayed with the Isaac Newton Institute for Mathematical Sciences, University of Cambridge, Cambridge, U.K., and HKBU for short visits.

His research interests include data fusion, image processing, deep learning, variational modeling and algorithms, and numerical PDEs.
\end{IEEEbiography}

\begin{IEEEbiography}[{\includegraphics[width=1in,height=1.25in,clip,keepaspectratio]{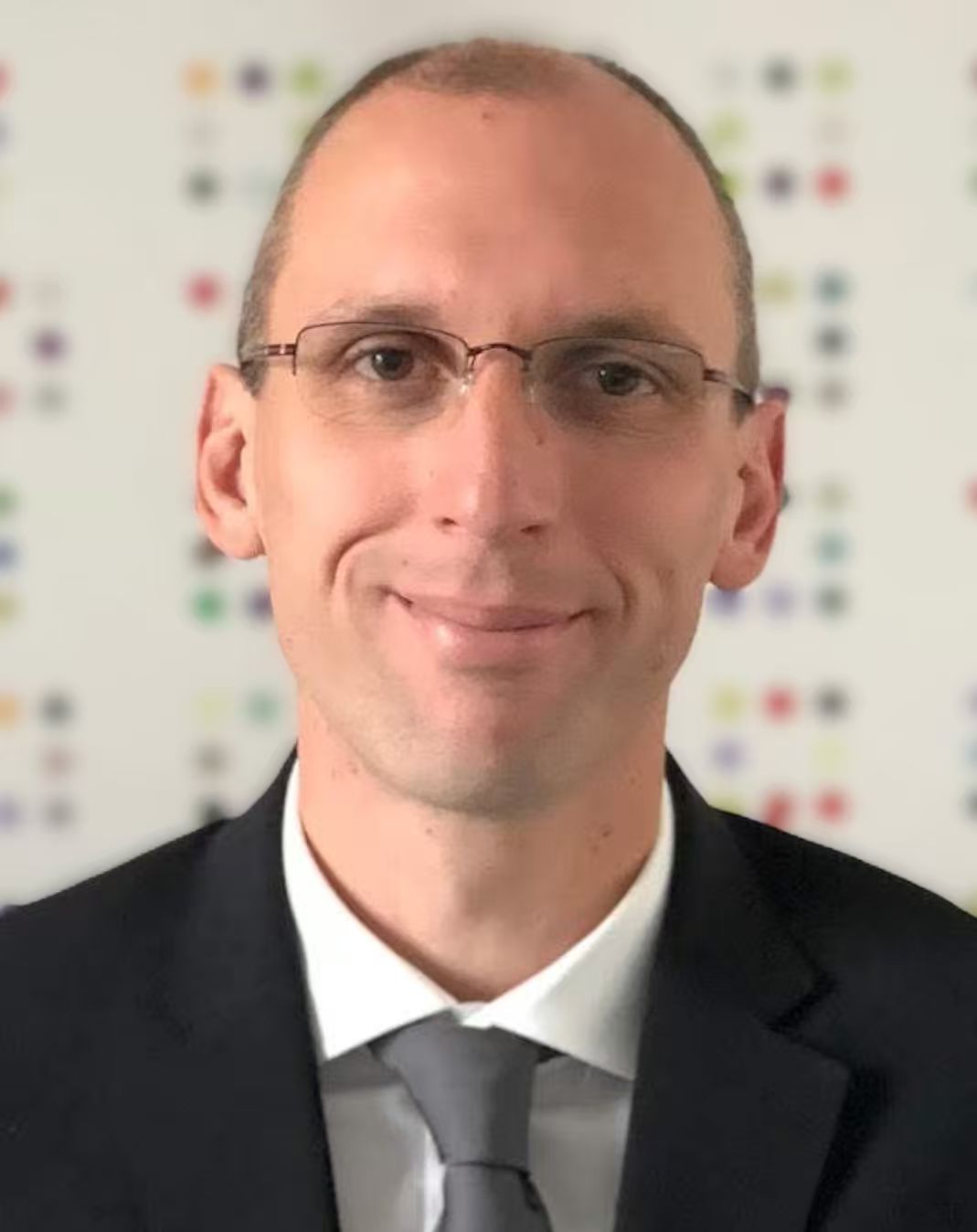}}]{Matja\v{z} Perc}
(Member, IEEE) received the Ph.D. degree in physics from the University of Maribor in 2006. He is currently a Professor in physics with the University of Maribor, a Staff Researcher with the Community Healthcare Center Dr. Adolf Drolc Maribor, and an Adjunct Professor with Kyung Hee University and Korea University.

He is a member of Academia Europaea and the European Academy of Sciences and Arts, and was among the top 1\% most cited physicists according to Clarivate Analytics data from 2020 to 2023. In 2019, he became a Fellow of the American Physical Society. He is also the 2015 recipient of the Young Scientist Award for Socio and Econophysics from the German Physical Society and the 2017 USERN Laureate. In 2018, he received the Zois Award, which is the highest national research award in Slovenia.
\end{IEEEbiography}

\begin{IEEEbiography}[{\includegraphics[width=1in,height=1.25in,clip,keepaspectratio]{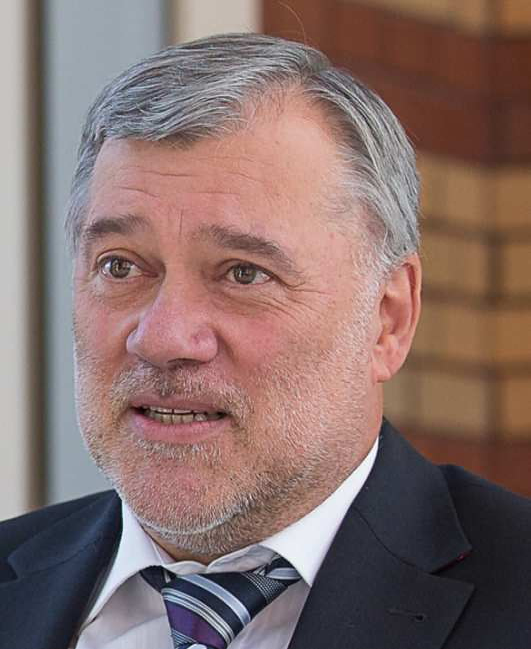}}]{J{\"u}rgen Kurths}
received the B.S. degree in mathematics from the University of Rostock, Rostock, Germany, the Ph.D. degree from the Academy of Sciences, German Democratic Republic, Berlin, Germany, in 1983, the Honorary degree from N. I. Lobachevsky State University, Nizhny Novgorod, Russia, in 2008, and the Honorary degree from Saratov State University, Saratov, Russia, in 2012.

From 1994 to 2008, he was a Full Professor with the University of Potsdam, Potsdam, Germany. Since 2008, he has been a Professor in nonlinear dynamics with theHumboldt University of Berlin, Berlin, Germany, and the Chair of the Research Domain Complexity Science with Potsdam Institute for Climate Impact Research, Potsdam, Germany.

He has authored more than 700 papers, which have been cited more than 60,000 times, with an h-index of 111. He is a Highly Cited Researcher in Engineering. His main research interests include synchronization, complex networks, and time series analysis and their applications.

He is a member of Academia Europaea. He is a Fellow of the American Physical Society, the Royal Society of Edinburgh, and the Network Science Society. He was a recipient of the Alexander von Humboldt Research Award from India in 2005 and from Poland in 2021, the Richardson Medal of the European Geophysical Union in 2013, and the Eight Honorary Doctorates. He is the Editor-in-Chief of Chaos and serves on the editorial boards of more than ten journals.
\end{IEEEbiography}

\end{document}